\pdfoutput=1
\RequirePackage{ifpdf}
\ifpdf % We are running pdfTeX in pdf mode
\documentclass[pdftex]{sigma}
\else
\documentclass{sigma}
\fi

\newcommand{\C}{\mathbb{C}}

\newcommand{\la}{\lambda}

\newcommand{\Z}{\mathbb{Z}}
\newcommand{\R}{\mathbb{R}}

\begin{document}

\allowdisplaybreaks

\renewcommand{\thefootnote}{$\star$}

\renewcommand{\PaperNumber}{043}

\FirstPageHeading

\ShortArticleName{Darboux Integrals for Schr\"odinger Planar Vector Fields via Darboux Transformations}

\ArticleName{Darboux Integrals for Schr\"odinger Planar\\
 Vector Fields via Darboux Transformations\footnote{This
paper is a contribution to the Special Issue ``Geometrical Methods in Mathematical Physics''. The full collection is available at \href{http://www.emis.de/journals/SIGMA/GMMP2012.html}{http://www.emis.de/journals/SIGMA/GMMP2012.html}}}

\Author{Primitivo B.~ACOSTA-HUM\'ANEZ~$^\dag$ and Chara PANTAZI~$^\ddag$}

\AuthorNameForHeading{P.B.~Acosta-Hum\'anez and Ch.~Pantazi}

\Address{$^\dag$~Departamento de Matem\'aticas y Estad\'istica
Universidad del Norte,\\
\hphantom{$^\dag$}~Km.~5 via Puerto Colombia, Barranquilla, Colombia}
\EmailD{\href{mailto:pacostahumanez@uninorte.edu.co}{pacostahumanez@uninorte.edu.co}}

\Address{$^\ddag$~Departament de Matem\`atica Aplicada I, Universitat Polit\`ecnica de Cata\-lunya, (EPSEB),\\
\hphantom{$^\ddag$}~Av. Doctor Mara\~{n}\'on, 44--50, 08028 Barcelona, Spain}
\EmailD{\href{mailto:chara.pantazi@upc.edu}{chara.pantazi@upc.edu}}

\ArticleDates{Received March 05, 2012, in f\/inal form July 06, 2012; Published online July 14, 2012}

\Abstract{In this paper we study the Darboux transformations of planar vector f\/ields of Schr\"odinger type. Using the isogaloisian property of Darboux transformation we prove the ``invariance'' of the objects of the ``Darboux theory of integrability''. In particular, we also show how the shape invariance property of the potential is important in order to preserve the structure of the
 transformed vector f\/ield. Finally, as illustration of these results, some examples of planar vector f\/ields coming from supersymmetric quantum
  mechanics are studied.}

\Keywords{Darboux theory of integrability; Darboux transformations; dif\/ferential Galois theory; Schr\"odinger equation; supersymmetric quantum mechanics}

\Classification{12H05; 34A30; 34C14; 81Q60; 32S65}

\renewcommand{\thefootnote}{\arabic{footnote}}
\setcounter{footnote}{0}

\section{Introduction}
We deal with  some generalization of the  Darboux theory
of integrability for planar vector f\/ields
and the Darboux transformation of the associated equation.

In 1882, Darboux in his paper~\cite{da1}  presented  as a proposition the notable
theorem today known as \emph{Darboux transformation}. This proposition also
can be found in his book \cite[p.~210]{da2}. Curiously Darboux's proposition was forgotten for a long time.
In 1926 Ince included it in his book  as an exercise (see Exercises 5, 6 and 7 in \cite[p.~132]{in}). Ince follows closely
Darboux's formulation given in \cite{da1,da2}.  P.~Dirac, in 1930,
published \textit{The principles of quantum mechanics},  where he
gave a mathematically rigorous formulation of quantum mechanics. In
1938, J.~Delsarte introduced the notion of transformation
(transmutation) operator, today known as \emph{intertwining operator}
and is closely related with Darboux transformation and ladder
operators.  Later on, in 1941, E.~Schr\"odinger factorized in several ways the
hypergeometric equation. It was a byproduct of his
\textit{factorization method} originating an approach that can be
traced back to Dirac's raising and lowering operators for the
harmonic oscillator. Ten years later, in 1951, another factorization
method was presented by L.~Infeld and T.E.~Hull where they gave
the classif\/ication of their factorizations of linear second order
dif\/ferential equations for eigenvalue problems of wave mechanics. In
1955, M.M.~Crum inspired by  Liouville's work about Sturm--Liouville
systems and developed one kind of iterative generalization of
Darboux transformation. Crum surprisingly did not mention  Darboux.
In 1971, G.A.~Natanzon studied a general form of the transformation
that converts the hypergeometric equation to the Schr\"odinger
equation \mbox{writing} down the most general \emph{solvable potential},
potential for which the Schr\"odinger equation can be reduced to
hypergeometric or conf\/luent hypergeometric form, concept introduced
by himself.

Almost one hundred years  after  Darboux's proposition, in 1981,
Edward Witten with his renowned paper \cite{wi} gave birth to the
\emph{supersymmetric quantum mechanics} and  he discussed general
conditions for dynamical supersymmetry breaking. Since  Witten's
work we can found in the literature a big amount  of papers related to
supersymmetric quantum mechanics. Maybe the most relevant of these papers
was written in 1983 by L.\'E.~Gendenshtein, where  he presented the
\emph{shape invariance} condition, i.e.\ the preservation of the
shape under Darboux transformation.  Gendenshtein used this property to f\/ind the complete
spectra for a broad class of problems including all known exactly
solvable problems of quantum mechanics (bound state and
ref\/lectionless potentials). Today this kind of exactly solvable
potentials satisfying the shape invariance condition are called
\emph{shape invariant potentials}, see~\cite{ge}. In 2009, in
\cite{acthesis,acmowe} were presented a~Galoisian point of view of
the supersymmetric quantum mechanics and in particular of the
Darboux transformations and shape invariance condition. We point out that the f\/irst analysis of Darboux
transformations from the dif\/ferential Galois point of view was done by V.P.~Spiridonov in~\cite{spiridonov}. The present
work follows these approaches with the same point of view.

From the other hand, Darboux in 1878
 presented a simple way to construct f\/irst integrals and
integrating factors for planar polynomial vector f\/ields, see~\cite{Da}. The key point of his method are the invariant algebraic
curves of such vector f\/ields.  His approach has been related with
problems concering limit cycles, centers and bifurcation problems,
see for instance \cite{GL,LR,Sc}. Moreover, the geometric scenario
of the algebraic curves determines the structure of the vector
f\/ields, see \cite{CLPW2,CLPW, Pan}. Nowdays Darboux's method has been
improved for polynomial vector f\/ields basically taking into account
the multiplicity of the invariant algebraic curves see for instance \cite{CL1, CLPW,CLP}.
Moreover, the existence of a rational f\/irst integral and Darboux's
method are related by Jouanolou's results, see \cite{Jo,LLZ1}.  Prelle and Singer \cite{PS,Si}  gave the relation between
elementary/Liouvillian  f\/irst integrals and integrating factors that
are constructed by Darboux's method. Additionally, Darboux's ideas have
been extended   to a particular class of non-autonomous vector
 f\/ields, see \cite{BP1, LlPantazi}.

Our main aim in this paper is to relate Darboux's theory of
integrability for planar vector f\/ields of Riccati type  with Schr\"odinger equation
and Darboux transformation of the associated equation. So we consider polynomial vector f\/ields of the form
\begin{gather*}
\dot{v} = {dv}/{dt}=S_0(x)+S_1(x)v+S_2(x)v^2,\qquad
\dot{x} = {dx}/{dt}=N(x),
\end{gather*}
with $S_0,S_1,S_2\in\C[x].$ In particular for non-relativistic
quantum mechanics we have $S_2(x)=-N(x), S_1(x)=0$ and
$S_0(x)=N(x)(V(x)-\lambda)$  with $V(x)=T(x)/N(x),$  $N,T\in\C[x]$
and $\lambda$ a~constant.

 Hence, we deal with systems of the form
\begin{gather}
\dot{v}={dv}/{dt} = N(x)\left(V(x)-\lambda -v^2\right),\qquad
\dot{x}={dx}/{dt} = N(x),\label{pol1}
\end{gather}
or equivalently we can consider the polynomial vector f\/ield
\begin{gather}
\dot{v} = {dv}/{dt}=T(x)-N(x)\lambda -N(x)v^2,\qquad
\dot{x} = {dx}/{dt}=N(x), \label{vpol2}
\end{gather}
of degree  $m=\max\{\deg T(x),   \deg N(x)+2\}.$
Note that the associated foliation of system~\eqref{pol1} is
\begin{gather}\label{pol2}
v'=\dfrac{dv}{dx}=V(x)-\lambda -v^2,
\end{gather}
with $V\in\C(x)$, see also \cite{almp}.

The structure of the paper is the following: In Section~\ref{founds}
we present the basic concepts of dif\/ferential Galois theory,
Schr\"odinger equation, Darboux transformation and Darboux's theory
of integrability of planar polynomial vector f\/ields. In Section~\ref{mains} we present our main results. More concrete, in
Proposition~\ref{newprop} we characterize the dif\/ferential Galois
group of the Schr\"odinger equation with the class of the f\/irst
integral of the corresponding vector f\/ield. Additionally, in
Proposition~\ref{nonrational} we present a condition for the
non-existence of a rational f\/irst integral.  In Proposition~\ref{propdarp} we show how we can construct the generalized Darboux
f\/irst integrals and integrating factors of the transformed vector
f\/ield using a solution of the initial Schr\"odinger equation. In
Theorem~\ref{theo1} we show that the strong isogaloisian property of
the Darboux transformation and the shape invariance property of the
potential are necessary in order to preserve the rational structure
of the elements of the transformed vector f\/ield.
 At the end, in Section~\ref{sexamples} we give several examples as
applications  of our results.

As far as we know this is the f\/irst time that is presented in the literature
the link between Darboux theory of integrability of planar vector f\/ields
and  Darboux transformation of the associated Schr\"odinger equation.

\section{Theoretical background}\label{founds}

In this section we present the theoretical background that we use in this work.
 Most of the results are naturally extended in higher dimension.

\subsection{Dif\/ferential Galois theory}
We start regarding an algebraic model for functions and the corresponding Galois theory known as \emph{differential Galois theory} or also \emph{Picard--Vessiot theory}, see \cite{ka,kol,martinetramis,vasi,we2} for all detail.
 The following preliminaries correspond to a quick overview of this theory and can be found also in~\cite{acthesis, acmowe}.
\begin{definition}[dif\/ferential f\/ields]
Let $F$ be a commutative f\/ield of characteristic zero. A~{\em
derivation} of $F$ is a map $\frac{d}{dx} : F\rightarrow F$
satisfying
\[
\frac{d}{dx} (a + b) = \frac{d a}{dx}  + \frac{d b}{dx}, \qquad
\frac{d}{dx}(a\cdot b) = \frac{d a}{dx}  \cdot b + a \cdot\frac{d b}{dx},
\]
for all $a,b\in F$.  We say that $(F,\frac{d}{d{x}})$ (or just
$F$ when there is no ambiguity) is
a {\it{differential field}} with the derivation $\frac{d}{dx}$.

We assume that $F$ contains an element $x$ such that $\frac{d
}{d{x}}(x)=1$. Let $\mathcal C$ be the f\/ield of constants of $F$: $
\mathcal C = \left\{c\in F \,| \, \frac{d c}{dx}  = 0\right\} $.
$\mathcal C$ is of characteristic zero and will be assumed to be
algebraically closed.
\end{definition}

Throughout this paper, the \emph{coefficient field} for a
dif\/ferential equation will be def\/ined as the smallest dif\/ferential
f\/ield containing all the coef\/f\/icients of the equation.

 In
particular we deal with  second order linear homogeneous
dif\/ferential equations, i.e., equations of the form
\begin{gather}\label{soldeq}
\frac{d^2 y}{dx^2}+\alpha\frac{d y}{dx} +\beta y=0,\qquad \alpha,\beta\in F.
\end{gather}
\begin{definition}[Picard--Vessiot extension]
Consider the dif\/ferential equation \eqref{soldeq}. Let $L$ be a
dif\/ferential f\/ield containing $F$ (a dif\/ferential extension of $F$).
We say that~$L$ is a \emph{Picard--Vessiot} extension of $F$ for the
dif\/ferential equation~\eqref{soldeq} if there exist two linearly
independent  solutions of the dif\/ferential equation~\eqref{soldeq}
namely $y_1, y_2\in L$ such that $L= F\langle y_1, y_2 \rangle$ (i.e.\
$L=F(y_1, y_2,{dy_1}/{dx}, {dy_2}/{dx})$) and  moreover $L$ and $F$
have the same f\/ield of constants~$\mathcal{C}$.
\end{definition}

In what follows, we work with Picard--Vessiot extensions and the term
``solution of \eqref{soldeq}'' will mean ``solution of
\eqref{soldeq} in~$L$''.  So any solution of the dif\/ferential
equation~\eqref{soldeq} is a linear combination (over $\mathcal{C}$)
of~$y_{1}$ and~$y_{2}$.
\begin{definition}[dif\/ferential Galois groups]
An $F$-automorphism $\sigma$ of the Picard--Vessiot extension $L$ is called a dif\/ferential automorphism
if
\[
\sigma\left(\frac{da}{dx}\right)=\frac{d}{dx}(\sigma(a)) \quad
\forall \, a\in L \qquad \mbox{and} \qquad
 \sigma(a)=a   \quad  \forall\, a\in F.
\]
 The group of all dif\/ferential automorphisms of~$L$
over~$F$ is called the {\it differential Galois group} of~$L$
over~$F$ and is denoted by ${\rm DGal}(L/F)$.
\end{definition}

Given $\sigma \in \mathrm{DGal}(L/F)$, we see that $\{\sigma y_1,
\sigma y_2\}$ are also solutions of the equation \eqref{soldeq}.
Hence there exists a matrix $A_\sigma
\in \mathrm{GL}(2,\mathbb{C}),$ such that
\[
\sigma(
\begin{pmatrix}
y_{1} &
y_{2}
\end{pmatrix})
=
\begin{pmatrix}
\sigma (y_{1})  &
\sigma (y_{2})
\end{pmatrix}
=\begin{pmatrix} y_{1}& y_{2}
\end{pmatrix}A_\sigma.
\]
As $\sigma$ commutes with the derivation, this extends naturally to
an action on a fundamental solution matrix of the companion f\/irst
order system associated with the equation~\eqref{soldeq}. We have
\[\sigma \left(
\begin{pmatrix}
y_{1}&y_2\vspace{1mm}\\
\dfrac{d y_1}{dx}&\dfrac{d y_2}{dx}
\end{pmatrix} \right)
=
\begin{pmatrix}
\sigma (y_{1})&\sigma (y_2)\vspace{1mm}\\
\sigma \left(\dfrac{d y_1}{dx}\right)&\sigma \left(\dfrac{d y_2}{dx}\right)
\end{pmatrix}
=\begin{pmatrix} y_{1}& y_{2}\vspace{1mm}\\
\dfrac{d y_1}{dx}&\dfrac{d y_2}{dx}
\end{pmatrix}A_\sigma.
\]
This def\/ines a faithful representation $\mathrm{DGal}(L/K)\to
\mathrm{GL}(2,\mathbb{C})$ and it is possible to consider
$\mathrm{DGal}(L/K)$ as a subgroup of $\mathrm{GL}(2,\mathbb{C})$
and depends on the choice of the fundamental system $\{y_1,y_2\}$
 only up to conjugacy.

Recall that an algebraic group $G$ is an algebraic manifold endowed
with a group structure. Let $\mathrm{GL}(2,\mathbb{C})$ denote, as
usual, the set of invertible $2\times 2$  matrices with entries in
$\mathbb{C}$ (and $\mathrm{SL}(2,\mathbb{C})$ be the set of matrices
with determinant equal to $1$). A~linear algebraic group will be a
subgroup of $\mathrm{GL}(2,\mathbb{C})$ equipped with a structure of
algebraic group. One of the fundamental results of the
Picard--Vessiot theory is the following theorem (see~\cite{ka,kol}).

\begin{theorem}
The differential Galois group $\mathrm{DGal}(L/F)$ is an
algebraic subgroup of $\mathrm{GL}(2,\mathbb{C})$.
\end{theorem}
In fact, the dif\/ferential Galois group measures the algebraic
relations between the solutions (and their derivatives) of the
dif\/ferential equation \eqref{soldeq}.
It is sometimes viewed as the object which should tell ``what algebra sees of the dynamics of the solutions''.

In an algebraic group $G$, the largest connected algebraic subgroup
of $G$ containing the identity, noted $G^{\circ}$, is a normal
subgroup of f\/inite index. It is often called the {\it connected
component of the identity}. If $G=G^0$ then $G$ is a
\textit{connected group}.

When $G^0$ satisf\/ies some property,    we say that $G$ virtually
satisf\/ies this property. For example, {\it virtually solvability} of $G$
means solvability of $G^0$ (see~\cite{we2}).
\begin{theorem}[Lie--Kolchin] Let $G\subseteq
\mathrm{GL}(2,\mathbb{C})$ be a virtually solvable group. Then
$G^0$ is triangularizable, i.e.\ it is conjugate to a subgroup of
upper triangular matrices.
\end{theorem}

 Throughout this work we will use the following def\/inition.
\begin{definition}[Liouvillian integrability]\label{defint} We say that the linear
dif\/ferential equation \eqref{soldeq} is (Liouville)
\textit{integrable} if the Picard--Vessiot extension $L\supset F$ is
obtained as a tower of dif\/ferential f\/ields $F=L_0\subset
L_1\subset\cdots\subset L_m=L$ such that $L_i=L_{i-1}(\eta)$ for
$i=1,\ldots,m$, where either
\begin{enumerate}\itemsep=0pt
\item[1)] $\eta$ is {\emph{algebraic}} over $L_{i-1}$, that is $\eta$ satisf\/ies a
polynomial equation with coef\/f\/icients in $L_{i-1}$;
\item[2)] $\eta$ is {\emph{primitive}} over $L_{i-1}$, that is $\frac{d\eta}{dx} \in L_{i-1}$;
\item[3)] $\eta$ is {\emph{exponential}} over $L_{i-1}$, that is $\left(\frac{d\eta}{dx}\right)/\eta \in L_{i-1}$.
\end{enumerate}
\end{definition}

We remark that in the usual terminology of dif\/ferential algebra for
integrable equations  the corresponding Picard--Vessiot extensions
are called \textit{Liouvillian}. From now on we say that an equation
is {\it integrable} whether it is integrable in the sense of dif\/ferential
Galois theory according to Def\/inition~\ref{defint}.
The following theorem is due to Kolchin.
\begin{theorem}\label{LK} The equation \eqref{soldeq} is integrable if and only if
$\mathrm{DGal}(L/F)$ is virtually solvable.
\end{theorem}

There is an algorithm due to Kovacic~\cite{ko} that decides about the integrability of the equation~\eqref{soldeq} in the case where $F=\mathbb{C}(x)$.
In practice, Kovacic's algorithm deals with the reduced form of equation~\eqref{soldeq}, namely with the form $y''=ry$, where $r$ is a rational function.
 Kovacic used the fact that $\mathrm{DGal}(L/F)\subseteq\mathrm{SL}(2,\mathbb{C})$, in order to separate his algorithm in three cases for the integrability
 of the equation~\eqref{soldeq}:
\begin{description}\itemsep=0pt
\item[Case 1.] $\mathrm{DGal}(L/F)$ is reducible,
\item[Case 2.] $\mathrm{DGal}(L/F)$ is irreducible,
\item[Case 3.] $\mathrm{DGal}(L/F)$ is f\/inite primitive.
\end{description}

\subsection{The Schr\"odinger equation}
Here we f\/irst introduce the Schr\"odinger equation and then we present the preliminaries  about Schr\"odinger equation from   a Galoisian point of view,
see \cite{acthesis, acmowe}.

In classical mechanics  for a particle of mass $m$ moving under the
action of a potential $U$ the Hamiltonian is given by
\[
H={\|\vec{p}\|^2\over 2m}+U(\vec{x}),\qquad \vec{p}=(p_1,\ldots,p_n),\qquad \vec{x}=(x_1,\ldots,x_n),
\]
and corresponds to the energy (kinetic plus potential). From the
other hand in quantum mechanics the momentum $\vec{p}$ is
given by $\vec{p}=-\imath\hbar\nabla$, where $\hbar$ is
the Planck constant and $\nabla$ is the Laplacian operator. In this
case the Hamiltonian operator $H$ is the Schr\"odinger
(non-relativistic, stationary) operator which is given by
\[
H=-{\hbar^2\over 2m}\nabla^2+V(\vec{x}),
\]
where
 $\vec{x}$ is the {\it{coordinate}} and $V(\vec{x})$ is the {\it{potential
or potential energy}}. The Schr\"odinger equation is given by $H\Psi=\lambda \Psi$,  where the
eigenfunction $\Psi$ is the {\it{wave function}} and  the eigenvalue
$\lambda\in\mathrm{Spec} (H)$ is the {\it{energy level}}.  The solutions $\Psi$
of the Schr\"odinger equation are the {\it{states}} of the particle and $\mathrm{Spec} (H)$ denotes the \textit{spectrum} of the operator $H$. In~\cite{te} it can be found
the details about the mathematical foundations of quantum mechanics  for the Schr\"odinger equation.

According to \cite{cokasu2,wi}, a {\it supersymmetric quantum mechanical system}
is one in which there are operators $Q_i$ that commute with the Hamiltonian $\mathcal H$ and
satisfying
\begin{gather*}
[Q_i,\mathcal H]=Q_iH-HQ_i=0,\\
\{Q_i,Q_j\}=Q_iQ_j+Q_jQ_i=\delta_{ij}\mathcal H \qquad {\mbox{and}}
\qquad  \delta_{ij}=
\begin{cases}
1, & i=j,\\
0, & i\neq j.
\end{cases}
\end{gather*}
For $n = 2$, we obtain the simplest example of a supersymmetric quantum mechanical system.
In this case we have that $x\in\R$. Thus, the {\it supercharges}~$Q_i$
are def\/ined as{\samepage
\[
Q_\pm=\frac{\sigma_1p\pm\sigma_2W(x)}2,\qquad Q_+=Q_1,\qquad Q_-=
Q_2,
\]
where $p = -i\hbar\frac{d}{dx}$, $W:\R\longrightarrow \R$ is the {\it
superpotential} and $\sigma_i$ are the {\it Pauli spin matrices}.}

 The operator $\mathcal H$, satisfying $Q_i\mathcal
H=\mathcal HQ_i$ and $2Q_i^2=\mathcal H$, is given by
\[
 \mathcal H=\frac{I_2p^2+I_2W^2(x)+\hbar\sigma_3\frac{d}{dx}W(x)}2=
 \begin{pmatrix} H_+&0\\ 0&H_-
\end{pmatrix},\qquad I_2=
\begin{pmatrix}
1 & 0\\
0 & 1
\end{pmatrix}.
\]
The operators $H_-$ and $H_+$ are the {\it supersymmetric partner
Hamiltonians} and are given by
\begin{gather}\label{partner}
H_{\pm}=-\dfrac{1}{2}\frac{d^2}{dx^2}+V_{\pm}, \qquad
V_{\pm}=\left(\dfrac{W}{\sqrt{2}}\right)^2\pm \dfrac{1}{\sqrt{2}}\frac{d}{dx}\left(\dfrac{W}{\sqrt{2}}\right),
\end{gather}
where $V_\pm$ are the {\it supersymmetric partner potentials}.

Now we follow \cite{acthesis,acmowe} in order to present the
Schr\"odinger equation in the context of dif\/ferential Galois theory.
Thus, the Schr\"odinger equation (stationary and one dimensional) now
is written~as
\begin{gather}\label{equ1}
H\Psi=\lambda\Psi, \qquad
H=-\frac{d^2}{dx^2}+V(x),\qquad V\in F,
\end{gather}
where $F$ is a dif\/ferential f\/ield (with $\mathcal{C}=\mathbb{C}$ as f\/ield of
constants). We will deal with the integrability of equation~\eqref{equ1} in agreement with our def\/inition
of \emph{integrability}, i.e., in the sense of dif\/ferential Galois theory, see Def\/inition~\ref{defint}.
In \cite{acthesis,acmowe} were introduced the following notations, useful for our purposes.
\begin{itemize}\itemsep=0pt
\item $\Lambda\subseteq{\mathbb{C}}$ denotes the \textit{algebraic spectrum} of $H$, i.e., the set of
eigenvalues $\lambda$ such that equation~(\ref{equ1}) is
integrable (Def\/inition~\ref{defint}).
\item $L_\lambda$ denotes the Picard--Vessiot extension of equation~\eqref{equ1}. Thus, the dif\/ferential Galois group of \eqref{equ1} is denoted by
$\mathrm{DGal}(L_\lambda/K)$.
\end{itemize}

\begin{definition}[algebraically solvable and quasi-solvable potentials] The
potential $V(x)\in F$~is:
\begin{itemize}\itemsep=0pt
\item an \textit{algebraically solvable potential} when $\Lambda$ is an inf\/inite set, or
\item an \textit{algebraically quasi-solvable potential} when $\Lambda$ is a non-empty f\/inite
set, or
\item an \textit{algebraically non-solvable potential} when $\Lambda=\varnothing$.
\end{itemize}
When $\mathrm{Card}(\Lambda)=1$, we say that $V(x)\in F$ is a
\textit{trivial} algebraically quasi-solvable potential.
\end{definition}

The following theorem shows that if there exist more than one
 eigenvalue in the algebraic spectrum of the Schr\"odinger operator with $F=\mathbb{C}(x)$, then  we cannot fall in case~3 of Kovacic's algorithm.
\begin{theorem}[see \cite{acthesis,acmowe}]\label{lucky}
Consider the Schr\"odinger equation  \eqref{equ1} with
$F=\mathbb{C}(x)$ and Picard--Vessiot extension~$L_\lambda$.
 If $\mathrm{DGal}(L_0/F)$ is finite primitive, then $\mathrm{DGal}(L_\lambda/F)$ is not finite primitive for
  all $\lambda\in\Lambda\setminus \{0\}$.
\end{theorem}
From \cite{acthesis, acmowe} note that the known cases of rational
potentials in quantum mechanics leads to Schr\"odinger equations
falling in case~1 of Kovacic's algorithm. Additionally, if
$\mathrm{Card}(\Lambda)>1$ then any algebraic solution of the
Riccati equation associated to
 the Schr\"odinger equation~\eqref{equ1} is  a~root of a  polynomial of degree at most two.

\subsection{Darboux transformation}\label{darbouxsection}

Darboux gave in \cite{da1} a transformation that allow us to
transform some type of dif\/ferential equations into other
dif\/ferential equations preserving the type. The following results corresponds to the Darboux transformation, denoted as DT,
in the Galoisian and quantum mechanic formalism, see~\cite{acthesis,acmowe}.

\begin{theorem}[Galoisian version of DT]\label{darth} Assume $H_\pm=-\frac{d^2}{dx^2}+ V_{\pm}(x)$ and $\Lambda\neq\varnothing$.
Consider the Schr\"odinger equation $H_-\Psi^{(-)}=\lambda\Psi^{(-)}$
with $V_-(x)\in F$.
Let $\mathrm{DT}$ be the transformation
such that  $V_-\mapsto
V_{+}$,
 $\Psi^{(-)}\mapsto\Psi^{(+)}$, $F\mapsto \tilde{F}$.
  Then for the Schr\"odinger equation $H_+\Psi^{(+)}=\lambda\Psi^{(+)}$ with $V_+(x)\in\widetilde F$
 the following statements
 holds:
\begin{itemize}\itemsep=0pt
\item[$i)$] $\mathrm{DT}(V_-)=V_+=\Psi^{(-)}_{\lambda_1}\dfrac{d^2}{dx^2}\left(\dfrac{1}{\Psi^{(-)}_{\lambda_1}}\right)+\lambda_1=V_--2\displaystyle{d^2\over dx^2}\big(\ln\Psi^{(-)}_{\lambda_1}\big)$,\\
$\mathrm{DT}\big(\Psi^{(-)}_{\lambda_1}\big)={\Psi}^{(+)}_{\lambda_1}=\dfrac{1}{\Psi^{(-)}_{\lambda_1}}$,
 where\\
$\Psi^{(-)}_{\lambda_1}$ is a particular solution of $H_-{\Psi^{(-)}}=\lambda_1\Psi^{(-)}$, $\lambda_1\in\Lambda$;

\item[$ii)$] $\mathrm{DT}\big(\Psi^{(-)}_\lambda\big)=\Psi^{(+)}_{\lambda}=\dfrac{d}{dx}\Psi^{(-)}_{\lambda}-\dfrac{d}{dx}\big(\ln\Psi^{(-)}_{\lambda_1}\big)
\Psi^{(-)}_{\lambda}$, $\lambda\neq\lambda_1$, where

\noindent $\Psi^{(-)}_{\lambda}$ is the general solution of
$H_-{\Psi^{(-)}}=\lambda\Psi^{(-)}$ for
$\lambda\in\Lambda\setminus\{\lambda_1\}$ and

\noindent $\Psi^{(+)}_{\lambda}$ is the general solution of
$H_+{\Psi^{(+)}}=\lambda\Psi^{(+)}$ also for
$\lambda\in\Lambda\setminus\{\lambda_1\}$.
\end{itemize}
\end{theorem}

\begin{remark}\label{remarkiso}  According to \cite{acthesis,acmowe} we have that a transformation is called \emph{isogaloisian} whether
  it preserves the dif\/ferential Galois group: the initial equation and the transformed equation have the same dif\/ferential Galois group.
  Furthermore, when the dif\/ferential f\/ield and the Picard--Vessiot extension are preserved, then the transformation is called
  $\emph{strong isogaloisian}$.\end{remark}
  In agreement with  Theorem \ref{darth} and  Remark \ref{remarkiso}, we obtain the following
results, see \cite{acthesis,acmowe} for complete statements and proofs.

\begin{proposition}\label{darbiso} In general, $\mathrm{DT}$ is isogaloisian and virtually strong isogaloisian.
Furthermore, if $\partial_x(\ln\Psi^{(-)}_{\lambda_1})\in F$, then
$\mathrm{DT}$ is strong isogaloisian.
\end{proposition}

\begin{proposition}\label{sspp}
The supersymmetric partner potentials $V_\pm$ are rational
functions if and only if the superpotential $W$ is a rational
function.
\end{proposition}

\begin{corollary}\label{corsspp} The superpotential $W\in\mathbb{C}(x)$ if and only if $\mathrm{DT}$ is strong isogaloisian.
\end{corollary}
\begin{remark}
The examples of this paper are in agreement with
the previous results, since $\widetilde{F}=F$ due to the fact that
the superpotential $W$ belongs to  $F=\C(x)$. The following def\/inition is a~partial Galoisian adaptation of the original def\/inition given in~\cite{ge} ($F=\mathbb{C}(x)$). The complete Galoisian adaptation is
given when $F$ is any dif\/ferential f\/ield, see~\cite{acthesis,acmowe}.
\end{remark}

\begin{definition}[rational shape invariant potentials, see \cite{acthesis,acmowe}] Assume
$V_\pm(x;\mu)\in\mathbb{C}(x;\mu)$, where $\mu$ is a family of
parameters. The potential $V=V_-\in\mathbb{C}(x)$ is said to be
rational shape invariant potential with respect to $\mu$ and
$\lambda=\lambda_n$ being $n\in \mathbb{Z}_+$, if there exists a function $f$ such that
\begin{gather*}
V_+(x;a_0)=V_-(x;a_1)+R(a_1),\qquad a_1=f(a_0),\qquad
  \lambda_n=\sum_{k=2}^{n+1}R(a_k), \qquad  \lambda_0=0.
\end{gather*}
\end{definition}
Hence the form of the potentials $V_{\pm}$ are preserved up to parameters. Theorem \ref{darth} and Propositions~\ref{darbiso} and~\ref{sspp}
lead us to following result.

\begin{theorem}[see \cite{acthesis,acmowe}]\label{thsip} Consider $H\Psi^{(-)}=\lambda_n\Psi^{(-)}$ with Picard--Vessiot extension
$L_n$, where $n\in\mathbb{Z}_+$. If $V=V_-\in\mathbb{C}(x)$ is a
shape invariant potential with respect to $\lambda=\lambda_n$, then
\[
\mathrm{DGal}(L_{n+1}/\mathbb{C}(x))=\mathrm{DGal}(L_{n}/\mathbb{C}(x)),\qquad n>0.
\]
\end{theorem}

Hence for rational shape invariant potentials Galois group is preserved due to
Picard--Vessiot extension and dif\/ferential f\/ield.

\subsection{Darboux's theory of integrability for planar polynomial
vector f\/ields}

In this subsection we present the basic ideas of Darboux's method  for planar polynomial vector f\/ields, see~\cite{Da}. We don't give en extensive
presentation of this theory but we only present the basic results that we need in Section~\ref{mains}.

We consider the {\it polynomial $($differential$)$
system}  in $\C^2$ def\/ined by
\begin{gather}
\frac{dx}{dt}={\dot x}=P(x,y), \qquad   \frac{dy}{dt}={\dot
y}=Q(x,y), \label{1}
\end{gather}
 where $P$ and $Q$ are polynomials in the variables~$x$ and~$y$.
The independent variable~$t$ can be real or complex.
We associate to the polynomial dif\/ferential system~(\ref{1})
 the {\it polynomial vector field}
\begin{gather}
\label{2} X=P(x,y)\frac{\partial}{\partial x}+
Q(x,y)\frac{\partial}{\partial y},
\end{gather}
 and   its associated foliation is given by
$Pdy-Qdx=0.$

An algebraic curve $f(x,y)=0$ in $\C^2$ with $f \in \C[x,y]$ is an
{\it invariant algebraic curve} of the vector f\/ield (\ref{2}) if
\begin{gather}
X(\log(f))=K, \label{cofactor}
\end{gather}
for some polynomial $K\in \C[x,y]$ called the {\it cofactor}
of the invariant algebraic curve $f=0$. Note that due to the def\/inition \eqref{cofactor} we have that the degree of the cofactor $K$ is less than the
degree of the polynomial vector f\/ield \eqref{1}.
Moreover, the curve $f=0$ is
formed by trajectories of the vector f\/ield $X$.

For a given system (\ref{1}) of degree $m$ the computation of all
the invariant algebraic curves is a~very hard problem  because in
general we don't know about the maximum degree of such curves.
However, imposing additionally conditions either for the structure
of the system or for the nature of the curves we can have an
evidence of a such a bound~\cite{Car, CeLi, CLPZ}.

Let $h,g \in \C[x, y]$ be relatively prime in the ring $\C[x, y]$.
The function ${\exp}\left( g/h \right)$ is called an
{\it{exponential factor}} of the polynomial system (\ref{1}) if
there is  a polynomial $L\in \C[x,y]$ (also called cofactor)
that satisf\/ies the equation
$X\left( {g}/{h}\right)=L$.
It turns out that if $h$ is not a constant polynomial, then $h=0$ is
one of the invariant algebraic curve of \eqref{1}.

The following theorem is a short version of Darboux theory of
integrability for planar polynomial dif\/ferential systems. For more
details and also for higher dimension see \cite{LL}. For genera\-li\-zations see~\cite{BP1}.
\begin{theorem} \label{theodar}
Suppose that a  polynomial system \eqref{1} of degree  $m$ admits
\begin{itemize}\itemsep=0pt
\item $p$ irreducible invariant algebraic curves $f_i=0$ with cofactors $K_i$ for $i=1,\dots,p$,
\item $q$ exponential factors $F_j$ with cofactors $L_j$ for $j=1,\dots,q.$
\end{itemize}
Then the following statements hold:
\begin{itemize}\itemsep=0pt
\item[$(a)$] The function
\begin{gather}
f_1^{\la_1} \cdots f_p^{\la_p} F_1^{\mu_1} \cdots
F_q^{{\mu}_q}, \label{int}
\end{gather}
is a  first integral of \eqref{1} if and only if
$ \sum\limits _{i=1}^p \la_i K_i+ \sum\limits
_{i=1}^q \mu_i L_i=0.$

\item[$(b)$]
The function \eqref{int} is a $($Darboux$)$ integrating
factor of the vector field \eqref{2} if and only if
$
\sum\limits_{i=1}^p \la_i K_i+ \sum\limits_{i=1}^q {\mu}_i L_i+(P_x+Q_y)=0.
$
\end{itemize}
\end{theorem}

Integrating factors for planar systems can be thought as
parametrization of the independent variable (time)
 that yields to a divergence free system.

As we show in Theorem~\ref{theodar} for polynomial vector f\/ields
the invariant algebraic curves and (because of their multiplicity, see~\cite{CLP}) the exponential factors are the basic elements in order
to construct f\/irst integrals/integrating factors. As we will see the method of Theorem \ref{theodar}
also works for more general expressions of vector f\/ields, curves, exponential factors and cofactors, see also~\cite{BP1, GGG,GG}.

\begin{definition}[generalized exponential factor]\label{genexp}
Consider $S(x)\in\C(x)$. We def\/ine a \emph{generalized exponential factor}  of a polynomial vector f\/ield $X$ any
expression of the form $F=\exp\left({\int S(x)}\right)$  which satisf\/ies $X(F)=LF$ and $L$ is called  \emph{generalized cofactor}.
\end{definition}

\begin{definition}[generalized Darboux function]\label{gendar}We def\/ine {\it generalized Darboux function} any expression
of the form
\begin{gather}
 (y-S_1(x))^{\lambda_1}\cdots(y-S_p(x))^{\lambda_p}\exp\left({\displaystyle{\int} S(x)}\right) \nonumber\\
\qquad{}= (y-S_1(x))^{\lambda_1}\cdots(y-S_p(x))^{\lambda_p}\exp{(\tilde{S}(x))}  \prod(x-x_i)^{b_i} ,\label{genfun}
\end{gather}
with $S_i,S,\tilde{S},g\in\C(x)$, $x_i,b_i,\lambda_i\in\C$ and $ p\in\Z_{+}$, see also~\cite{Zol}.
\end{definition}

An integrating factor  of the form \eqref{genfun}
will be called {\it generalized Darboux integrating factor} and similarly a f\/irst integral of the form \eqref{genfun} will be called
 {\it generalized Darboux first integral}.
\begin{remark}\label{casos}
We consider the Schr\"odinger equation~\eqref{equ1}  with $F=\C(x)$ and the associated Ricatti equation~\eqref{pol2}.
We also consider $v_1(x)$, $v_2(x)$, $v_3(x)$ particular solutions of equation~\eqref{pol2} with $V\in\C(x)$ and we  write $V(x)=T(x)/N(x)$ with
$T,N\in\C[x]$.
Note that an associated polynomial vector f\/ield
to equation \eqref{pol2} can be  written into the  form~\eqref{vpol2}.

$(a)$ Following \cite{almp,ko,Zol} we can distinguish the following cases about the type of the f\/irst integrals of equation~\eqref{pol2} or equivalently
of the polynomial vector f\/ield~\eqref{vpol2}.

\emph{Case 1.} $(i)$ If  only $v_1(x)\in\C(x)$ then the vector f\/ield \eqref{vpol2} has a f\/irst integral of
{\it Darboux--Schwarz--Christoffel type}, namely a f\/irst integral of the form
\begin{gather*}
I(v,x)=\dfrac{1}{y-v_1(x)}\exp({g(x)})\prod(x-x_i)^{a_i}
  +\int\limits^{x}
\exp({g(u)})\prod(u-x_i)^{a_i-m_i}P(u)du,%\label{dsch}
\end{gather*}
with $P\in\C[x]$, $g\in\C(x)$, $x_i,a_i\in\C$ and $m_i\in\Z_{+}$, see also~\cite{Zol}.

$(ii)$ If both  $v_1(x), v_2(x)\in\C(x)$ then the vector f\/ield \eqref{vpol2} has a generalized Darboux
f\/irst integral of the form
\begin{gather}\label{intvpol}
I(v,x)=\dfrac{-v+v_2(x)}{-v+v_1(x)}\exp\left({\int(v_2(x)-v_1(x))dx}\right).
\end{gather}
Furthermore if the dif\/ferential Galois group of the  Schr\"odinger
equation \eqref{equ1} is a cyclic group of order $k$ then the $k$-th power of the
 f\/irst integral \eqref{intvpol} is rational, see also \cite{almp}.

\emph{Case 2.} If $v_1$ is a solution of a quadratic polynomial  then the vector f\/ield \eqref{vpol2} has a  f\/irst integral of hyperelliptic type.

 \emph{Case 3.} If all $v_1$, $v_2$, $v_3$ are algebraic over $\C(x)$ then the vector f\/ield \eqref{vpol2} has a rational f\/irst integral of the form
\[
I(v,x)=\dfrac{(v_2-v_1)(v_1-v)}{(v_3-v_1)(v_2-v)}.
\]

$(b)$ In general, knowing one algebraic solution $v_1(x)$ of equation~\eqref{pol2} we can obtain a second solution of~\eqref{pol2}, namely
\[
v_2(x)=v_1+ \frac{\exp\left({-2\int v_1 dx}\right)}
{\int \exp\left({-2\int v_1dx}\right)dx}.
\]
Then, the vector f\/ield \eqref{vpol2} has always a f\/irst integral of the form~\eqref{intvpol}
and can be rewritten either as  Darboux--Schwarz--Christof\/fel type  or as a generalized Darboux function or as a f\/irst integral of hyperelliptic type
or as a rational f\/irst integral.
\end{remark}

Additionally in \cite{almp} appears the following result about the existence of a rational f\/irst integral of a polynomial vector f\/ield whose foliation is of
Riccati type.
\begin{theorem}\label{ciclic}
The Galois group of the equation $y''=r(x)y$ with $r\in\C(x)$ is finite if and only if there exist a rational first integral for
the associated polynomial vector field of
the corresponding Riccati equation.
\end{theorem}
Note that the Schr\"odinger equation \eqref{equ1} can be always
written into the form $y''=r(x)y$ and so we can always apply Theorem~\ref{ciclic}.

\section{Main results}\label{mains}

From now on we consider the Schr\"odinger equation \eqref{equ1} with potential
$V=V_-$:
\begin{gather}
\Psi^{(-)\prime\prime}=(V_-(x)-\lambda)\Psi^{(-)},
 \qquad V_-=\dfrac{T}{N},\qquad T,N\in\mathbb{C}[x],\nonumber\\
  \lambda\in\Lambda, \qquad \mathrm{Card}(\Lambda)>1, \qquad F=\mathbb{C}(x).\label{estrella}
  \end{gather}
 After the change of coordinates $\zeta^{(-)}={\Psi^{(-)\prime}}/\Psi^{(-)}$,  equation~\eqref{equ1} can be written as
\begin{gather}
{\zeta^{(-)\prime}} = V_--\lambda -{\zeta^{(-)2}},
\label{partida}
\end{gather}
and we associate to equation \eqref{partida} the polynomial vector f\/ield
\begin{gather}
X^{(-)}_\lambda=\big(T-\lambda N-N{\zeta^{(-)2}}\big)\dfrac{\partial }{\partial \zeta^{(-)}}+N\dfrac{\partial }{\partial x},
\label{partidav}
\end{gather}
of degree $m=\max\{\deg T(x),  \deg N(x)+2\}.$
 By Propositions~\ref{darbiso}, \ref{sspp} and Corollary \ref{corsspp}, we have that for given $V_-\in F$ and $W=\zeta^{(-)}_0=\Psi^{(-)\prime}_0/\Psi^{(-)}_0\in F$
  then we obtain that $V_+\in F$, i.e., the
Darboux transformation $\mathrm{DT}$ is strong isogaloisian.
Moreover, the superpotential~$W$ is rational. The applications
considered in this work correspond to this case.

The following proposition follows directly by Remark~\ref{casos} and~\cite{almp}.

\begin{proposition}\label{newprop}
The differential Galois group $\mathrm{DGal}(L_\lambda/F)$ of the Schr\"odinger equation \eqref{estrella} is virtually solvable if and only if
the first integral of the vector field \eqref{partidav}
can be written in one of the forms appearing in Remark~{\rm \ref{casos}$(a)$}.
\end{proposition}

In  the next proposition we present a result about the non-existence of a rational f\/irst integral.

\begin{proposition}\label{nonrational}
Consider  the Schr\"odinger equation \eqref{estrella}. If  $\mathrm{DGal}(L_\lambda/F)$ is not cyclic then the associated foliation
of the Schr\"odinger equation \eqref{estrella} has not rational first integrals.
\end{proposition}

\begin{proof}
It follows directly from Theorems \ref{lucky} and \ref{ciclic}.
\end{proof}

\begin{lemma}\label{lemdarp}
Consider $\Psi^{(-)}_\lambda(x)$ a solution of the Schr\"odinger equation \eqref{estrella}  and we denote by
\begin{gather}
\zeta^{(-)}_{\lambda}(x)=\dfrac{\big(\Psi_{\lambda}^{(-)}(x)\big)'}{\Psi^{(-)}_{\lambda}(x)}=
\big(\ln\big( \Psi_{\lambda}^{(-)}(x)\big)\big)',\qquad \lambda\in \Lambda.
\label{sc1}
\end{gather}
Then for all $\lambda\in \Lambda$ the vector field \eqref{partidav}
admits the following:
\begin{itemize}\itemsep=0pt
\item Invariant  curve
\[
f^{(-)}_\lambda\big(\zeta^{(-)},x\big)=-\zeta^{(-)}+\zeta^{(-)}_{\lambda}(x),
\]
with generalized cofactor
\[
K^{(-)}_\lambda\big(\zeta^{(-)},x\big)=-N(x)\big(\zeta^{(-)}+\zeta^{(-)}_{\lambda}(x)\big).
\]
\item Generalized exponential factor
\[
F^{(-)}_\lambda\big(\zeta^{(-)},x\big)=\exp\left(\displaystyle{\int}\left(\dfrac{1}{2}\dfrac{N'(x)}{N(x)}+ \zeta^{(-)}_{\lambda}(x)\right)dx\right),
\]
with generalized cofactor
\[
L^{(-)}_\lambda\big(\zeta^{(-)},x\big)={N'(x)}/{2}+N(x)\zeta^{(-)}_{\lambda}(x).
\]
\item An integrating factor of the form
\[
R^{(-)}_\lambda\big(\zeta^{(-)},x\big)=\dfrac{\exp\left({\displaystyle{\int}\left(-\dfrac{N'(x)}{N(x)}-2 \zeta^{(-)}_{\lambda}(x)\right)dx}\right)}{\big({-}\zeta^{(-)}+\zeta^{(-)}_{\lambda}(x)\big)^2} .
\]
\item A first integral of the form
\[
I_\lambda^{(-)}\big(\zeta^{(-)},x\big)=\dfrac{-\zeta^{(-)}+\zeta_{(\lambda,2)}^{(-)}}{-\zeta^{(-)}+\zeta_{(\lambda,1)}^{(-)}}
\exp\left({\displaystyle{\displaystyle{\int}\big( \zeta_{(\lambda,2)}^{(-)}-\zeta_{(\lambda,1)}^{(-)}\big)dx }}\right),
\]
with $\zeta_{(\lambda,1)}$ as in \eqref{sc1} and
\[
\zeta_{(\lambda,2)}^{(-)}=\zeta_{(\lambda,1)}^{(-)}
+\displaystyle{\dfrac{\exp\left({-2\displaystyle{\int} \zeta_{(\lambda,1)}^{(-)}}dx\right)}{\displaystyle{\int} \exp\left(
{-2\displaystyle{\int} \zeta_{(\lambda,1)}^{(-)}}dx\right)dx}}.
\]
\end{itemize}
\end{lemma}
\begin{proof}
Note that $\zeta^{(-)}_{\lambda}(x)$ is a solution of the associated
equation ${\zeta^{(-)}}'=V_-(x)-\lambda-{\zeta^{(-)}}^2$ for all
$\lambda\in\Lambda$. Using the expression of the vector f\/ield \eqref{partidav} we have
\begin{gather*}
X_\lambda^{(-)}\big(f^{(-)}_\lambda(\zeta^{(-)},x)\big) =\big(T-\lambda N-N\zeta^{(-)2}\big)(-1)+N\zeta_\lambda^{(-)\prime}\\
\hphantom{X_\lambda^{(-)}\big(f^{(-)}_\lambda(\zeta^{(-)},x)\big)}{}
=N\big(\zeta^{(-)2}-\zeta_\lambda^{(-)2}\big)
=K_\lambda^{(-)}\big(\zeta^{(-)},x\big)\cdot f_\lambda^{(-)}\big(\zeta^{(-)},x\big).
\end{gather*}
 Note that the curve
$f^{(-)}_\lambda$ is polynomial in the variable $\zeta^{(-)}$ but
it could
  be not polynomial in the variable~$x$.
 Direct computations shows that system~\eqref{partida} admits
 the generalized exponential factor
\[
F^{(-)}_\lambda\big(\zeta^{(-)},x\big)=\exp\left({\int \left(\frac{1}{2}\frac{N'(x)}{N(x)}+\zeta^{(-)}_{\lambda}(x)\right)dx}\right),
\]
 with generalized cofactor
$L^{(-)}_\lambda(\zeta^{(-)},x)={N'(x)}/{2}+\zeta^{(-)}_{\lambda}(x)$.
 Note that vector f\/ield~\eqref{partidav}
has divergence $\mbox{div}(X_\lambda)=-2N\zeta^{(-)}+N'(x)$. Hence, we have
that
$
-2K^{(-)}_\lambda-2L^{(-)}_\lambda+\mbox{div}\big(X^{(-)}_\lambda\big)=0
$
and similarly to Theorem~\ref{theodar}$(b)$ we have that
$
R^{(-)}_\lambda=\big({{f^{(-)}_\lambda}{F^{(-)}_\lambda}}\big)^{-2}
$
is an integrating factor for the vector f\/ield~\eqref{partidav}.

If $\Psi^{(-)}_{(\lambda,1)}(x)$ a solution of the Schr\"odinger equation \eqref{estrella} then
$\zeta_{(\lambda,1)}^{(-)}$ is a solution of the Riccati equation~\eqref{partida}.
Then, according to Remark \eqref{casos}$(b)$ we have that $\zeta_{(\lambda,2)}^{(-)}$ is another solution of the Riccati equation \eqref{partida}.
In this case the vector f\/ield \eqref{partidav} admits the two invariant curves
\[
f^{(-)}_{(\lambda,1)}\big(\zeta^{(-)},x\big)=-\zeta^{(-)}+\zeta^{(-)}_{(\lambda,1)}(x), \qquad
f^{(-)}_{(\lambda,2)}\big(\zeta^{(-)},x\big)=-\zeta^{(-)}+\zeta^{(-)}_{(\lambda,2)}(x),
\]
with generalized cofactors
\[
K^{(-)}_{(\lambda,1)}\big(\zeta^{(-)},x\big)=-N(x)\big(\zeta^{(-)}+\zeta^{(-)}_{(\lambda,1)}(x)\big), \qquad
K^{(-)}_{(\lambda,2)}\big(\zeta^{(-)},x\big)=-N(x)\big(\zeta^{(-)}+\zeta^{(-)}_{(\lambda,2)}(x)\big).
\]
The two generalized exponential factors
\begin{gather*}
F^{(-)}_{(\lambda,1)}\big(\zeta^{(-)},x\big)= \exp\left({\displaystyle{\int}\left(\dfrac{1}{2}\dfrac{N'(x)}{N(x)}+ \zeta^{(-)}_{\lambda,1}(x)\right)dx}\right),\\
F^{(-)}_{(\lambda,2)}\big(\zeta^{(-)},x\big)=\exp\left({\displaystyle{\int}\left(\dfrac{1}{2}\dfrac{N'(x)}{N(x)}+ \zeta^{(-)}_{\lambda,2}(x)\right)dx}\right).
\end{gather*}
with generalized cofactors
\begin{gather*}
L^{(-)}_{(\lambda,1)}\big(\zeta^{(-)},x\big) ={N'(x)}/{2}+N(x)\zeta^{(-)}_{\lambda,1}(x),\qquad
L^{(-)}_{(\lambda,2)}\big(\zeta^{(-)},x\big) ={N'(x)}/{2}+N(x)\zeta^{(-)}_{\lambda,2}(x).
\end{gather*}
Note that it holds
$
-K^{(-)}_{(\lambda,1)}+K^{(-)}_{(\lambda,2)}-L^{(-)}_{(\lambda,1)}+L^{(-)}_{(\lambda,2)}=0,
$
and similar to Theorem~\ref{theodar}$(a)$ we have that the vector f\/ield $X_{\lambda}^{(-)}$ admits the f\/irst integral
\[
I^{(-)}_{(\lambda)}=\dfrac{f^{(-)}_{(\lambda,2)}\cdot F^{(-)}_{(\lambda,2)}}{f^{(-)}_{(\lambda,1)}\cdot F^{(-)}_{(\lambda,1)}}.
\]
Hence, the lemma is proved.
\end{proof}

\begin{proposition}\label{propdarp}
Consider $\Psi^{(-)}_0(x)$ a particular solution of the Schr\"odinger equation $H_-\Psi^{(-)}=\lambda\Psi^{(-)}$ for $\lambda=0$ and let
$\zeta^{(-)}_{0}(x)=\big(\ln\big( \Psi_{0}^{(-)}(x)\big)\big)'$.
Then after the Darboux transformation  $DT$ for all $\lambda\neq 0$
equation \eqref{partida} becomes
\begin{gather}
{\zeta^{(+)\prime}} = V_+-\lambda-{\zeta^{(+)2}},
\label{llegada}
\end{gather}
with $V_+=-{{\zeta_0}^{(-)\prime}}+{{\zeta_0}^{(-)2}}$ and its
associated vector field is
\begin{gather}
X^{(+)}_\lambda=\big({-}\zeta_0^{(-)\prime}+{\zeta_0^{(-)2}}-\lambda-{\zeta^{(+)2}}\big)\dfrac{\partial }{\partial {\zeta^{(+)}}}+\dfrac{\partial }{\partial x}.
\label{llegadav}
\end{gather}
 The vector field \eqref{llegadav} for all $\lambda\neq 0$ admits the following:
\begin{itemize}\itemsep=0pt
 \item Invariant curve
\[
{f}_\lambda^{(+)}\big({\zeta^{(+)}}, x\big)
=-{\zeta}^{(+)}+\zeta_{\lambda}^{(-)}+\big(\ln\big(\zeta_{\lambda}^{(-)}-\zeta^{(-)}_0\big)\big)',
\]
with generalized cofactor
\[
{K}_\lambda^{(+)}\big({\zeta}^{(+)}, x\big)=
-{\zeta}^{(+)}-\zeta_{\lambda}^{(-)}-\big(\ln\big(\zeta_{\lambda}^{(-)}-\zeta^{(-)}_0\big)\big)'.
\]
\item Generalized exponential factor
\[
{F}_\lambda^{(+)}\big({\zeta}^{(+)},x\big)=
{\big(\zeta_{\lambda}^{(-)}-\zeta^{(-)}_0\big)}{\exp\left({\int\zeta_{\lambda}^{(-)}}\right)},
\]
with generalized cofactor
\[
{L}_\lambda^{(+)}= \zeta_{\lambda}^{(-)}+\big(\ln\big(\zeta_{\lambda}^{(-)}-\zeta^{(-)}_0\big)\big)'.
\]
\item  An integrating factor  $($for $\lambda\neq 0)$
\[
{R}_\lambda^{(+)}\big({\zeta^{(+)}},x\big)=\frac{\exp\left({-2\displaystyle{\int \zeta_\lambda^{(-)}}}\right)}{\big(\zeta_\lambda^{(-)}
-\zeta_0^{(-)}\big)^2\big({-}{\zeta^{(+)}}+\zeta_\lambda^{(-)}+\big(\ln (\zeta_\lambda^{(-)}-\zeta_0^{(-)})\big)^\prime\big)^2}.
\]
\item A first integral of the form
\begin{gather*}
I_\lambda^{(+)}\big(\zeta^{(+)},x\big) = \dfrac{-\zeta^{(+)}+\zeta_{(\lambda,2)}^{(-)}+\big(\ln\big(\zeta_{(\lambda,2)}^{(-)}-\zeta^{(-)}_0\big)\big)'}
{-\zeta^{(+)}+\zeta_{(\lambda,1)}^{(-)}+\big(\ln\big(\zeta_{(\lambda,1)}^{(-)}-\zeta^{(-)}_0\big)\big)'}\\
\hphantom{I_\lambda^{(+)}\big(\zeta^{(+)},x\big) =}{}
\times \left(\dfrac{\zeta_{(\lambda,2)}^{(-)}-\zeta_0^{(-)}}{\zeta_{(\lambda,1)}^{(-)}-\zeta_0^{(-)}}\right)
\exp\left({\displaystyle{\int \big(\zeta_{(\lambda,2)}^{(-)}-\zeta_{(\lambda,1)}^{(-)}\big)dx }}\right).
\end{gather*}
\end{itemize}
\end{proposition}

\begin{proof}
For $\lambda=0$  we consider  $\Psi_{0}^{(-)}$ a particular solution  of the  Schr\"odinger equation $H_-\Psi^{(-)}=\lambda\Psi^{(-)}$.
We denote by  $\zeta^{(-)}_0=\left(\ln\left(\Psi_0^-\right)\right)'$. Then
from Theorem \ref{darth} for $\lambda\neq 0$   we have
\[
\Psi_\lambda^{(+)}=\Psi^{(-)\prime}_\lambda-\zeta^{(-)}_0\Psi^{(-)}_\lambda= \exp\left({\int \zeta_{\lambda}^{(-)}}\right)\big(\zeta_{\lambda}^{(-)}-\zeta^{(-)}_0\big),
\]
and we have use that $\zeta_{\lambda}^{(-)}=\Psi_\lambda^{(-)'}/\Psi_\lambda^{(-)}.$
Note that $\Psi^{(+)}_\lambda$ is a solution  of the Schr\"odinger equation $H_+\Psi^{(+)}=\lambda\Psi^{(+)}$.
Let
\[
\zeta_\lambda^{(+)}=\big(\ln\big(\Psi_\lambda^{(+)}\big)\big)'=\left(\ln\left( \exp\left({\int \zeta_{\lambda}^{(-)}}\right)\big(\zeta_{\lambda}^{(-)}-\zeta^{(-)}_0\big)  \right)\right)'.
\]
According to Theorem \ref{darth} we have
\[
V _+=\Psi_0^{(-)}\left(\frac{1}{\Psi_0^{(-)}}\right)^{\prime\prime}=\Psi_0^{(-)}\left(-\frac{\zeta_0^{(-)}}{\Psi_0^{(-)}}\right)^\prime=-\zeta_0^{(-)\prime}
+{\zeta_0^{(-)2}},
\]
and  we additionally  consider the change of variables $\zeta^{(+)}=(\ln(\Psi^{(+)}))'$.
Then the Schr\"odinger equation $H_+\Psi^{(+)}=\lambda\Psi^{(+)}$
falls in equation \eqref{llegada}.

Therefore, similar to  Lemma \ref{lemdarp}, the vector f\/ield \eqref{llegadav} for all $\lambda\neq 0$ admits the invariant curve~${f}_\lambda^{(+)}$ with generalized  cofactor
${K}_\lambda^{(+)}$ given by
\begin{gather*}
{f}_\lambda^{(+)}\big({\zeta^{(+)}},  x\big)=-{\zeta^{(+)}}+{\zeta}_\lambda^{(+)}
 = -{\zeta}^{(+)}+\zeta_{\lambda}^{(-)}+\big(\ln\big(\zeta_{\lambda}^{(-)}-\zeta^{(-)}_0\big)\big)',\\
{K}_\lambda^{(+)}\big({\zeta}^{(+)}, x\big)= -{\zeta}^{(+)}-{\zeta}_\lambda^{(+)}
=-{\zeta}^{(+)}-\zeta_{\lambda}^{(-)}-\big(\ln\big(\zeta_{\lambda}^{(-)}-\zeta^{(-)}_0\big)\big)'.
\end{gather*}
Additionally, for all $\lambda\neq 0$ admits the generalized exponential factor
\begin{gather*}
{F}_\lambda^{(+)}\big({\zeta}^{(+)},x\big) = \exp\left({\displaystyle{\int \zeta_{\lambda}^{(+)}}}\right)
=\exp\left({\displaystyle{\int\big( \zeta_{\lambda}^{(-)}+\big(\ln\big(\zeta_{\lambda}^{(-)}-\zeta^{(-)}_0\big)\big)'\big)}}\right)\\
\hphantom{{F}_\lambda^{(+)}\big({\zeta}^{(+)},x\big)}{}
 = {\big(\zeta_{\lambda}^{(-)}-\zeta^{(-)}_0\big)}{\exp\left({\int\zeta_{\lambda}^{(-)}}\right)}.
\end{gather*}
with generalized cofactor
$
{L}_\lambda^{(+)}=\zeta_{\lambda}^{(+)}= \zeta_{\lambda}^{(-)}+\big(\ln\big(\zeta_{\lambda}^{(-)}-\zeta^{(-)}_0\big)\big)'.
$
Note that the vector f\/ield~\eqref{llegadav} has divergence $\mbox{div}{X}=-2{\zeta}^{(+)}$ and for all $\lambda\neq 0$ it holds
$
-2{K}_\lambda^{(+)}-2{L}_\lambda^{(+)}+\mbox{div}{X_\lambda}^{(+)}=0,
$
and similarly to Theorem~\ref{theodar}$(b)$ we have that
$
{R}_\lambda^{(+)}={{{f_\lambda}^{(+)}}{{F_\lambda}^{(+)}}}^{-2}.
$

Note that if $\Psi^{(+)}_{(\lambda,1)}$ is a solution  of the Schr\"odinger equation $H_+\Psi^{(+)}=\lambda\Psi^{(+)}$ then
$\zeta^{(+)}_{(\lambda,1)}$ is a solution of the Riccati equation $\zeta^{(+)\prime}_\lambda=V_{+}-\lambda-\zeta^{(+)2}$.
Following the same arguments as in the proof of Lemma~\ref{lemdarp}
we obtain the expression of the f\/irst integral.
Hence, the proof is completed.
\end{proof}

\begin{theorem}\label{theo1} If Darboux transformation is strong isogaloisian for a potential $V_-\in \mathbb{C}(x)$, then each pair of functions
 $f^{(\pm)}_\lambda(\zeta^{(\pm)},x)$, $K^{(\pm)}_\lambda(\zeta^{(\pm)},x)$, $L^{(\pm)}_\lambda(\zeta^{(\pm)},x)$ and $I^{(\pm)}_\lambda(\zeta^{(\pm)},x)$
 belong to the differential extension in the variables $\zeta^{(\pm)}$ and $x$.
  Furthermore, if $V_-$ is a rational shape invariant potential, then
  $f^{(\pm)}_\lambda(\zeta^{(\pm)},x)$, $K^{(\pm)}_\lambda(\zeta^{(\pm)},x)$ and $L^{(\pm)}_\lambda(\zeta^{(\pm)},x)$ are rational functions
  in~$\zeta^{(\pm)}$ and~$x$.
\end{theorem}

\begin{proof}
The proof follows directly by Theorem \ref{thsip} and Proposition \ref{propdarp}.
\end{proof}

\section{Applications  in quantum mechanics}\label{sexamples}

In this section we present some examples that relate polynomial
vector f\/ields, according to Lemma~\ref{lemdarp} and Proposition~\ref{propdarp}, with supersymmetric quantum mechanics. Schr\"odinger equations with
potentials such as free particle (explicitly discussed in~\cite{spiridonov} with a similar approach),  three dimensional harmonic oscillator
and Coulomb potential are analyzed as Schr\"odinger vector f\/ields.

\begin{example}[free particle]
Consider the Schr\"odinger equation
$H_-\Psi^{(-)}=\lambda\Psi^{(-)}$ with potential $V_-=0$ and
dif\/ferential f\/ield\footnote{Although the smallest dif\/ferential f\/ield
containing the coef\/f\/icients of the Schr\"odinger equation is
$\mathbb{C}$, to avoid triviality, we choose $\mathbb{C}(x)$ as the
suitable dif\/ferential f\/ield for such equation.} $F=\mathbb{C}(x)$.
Thus, we have that $\Lambda=\mathbb{C}$.

 If we
choose $\lambda_0=0$ and as particular solution of the Schr\"odinger
equation the solution $\Psi_{0}^{(-)}=x$ we have that
$\zeta_0^{(-)}=\Psi_{0}^{(-)'}/ \Psi_{0}^{(-)}=1/x$. Since
$\zeta_0^{(-)}\in\C(x)$ from Proposition \ref{darbiso} we have that
$DT$ is strong isogaloisian. Additionally,
\[
\mathrm{DT}(V_-)=V_+=-{\zeta_0^{(-)\prime}}+{\zeta_0^{(-)2}}={2\over x^2},
\]
and so $V_-$ is not shape invariant.
Then for
$\lambda\neq 0$ the general solution of
$H_-\Psi^{(-)}=\lambda\Psi^{(-)}$ is given by
\[
\Psi^{(-)}_{\lambda} =  c_1\Psi^{(-)}_{(\lambda,1)}+c_2\Psi^{(-)}_{(\lambda,2)},
\]
where we have consider $\Psi^{(-)}_{(\lambda,1)}=\exp\left({\sqrt{-\lambda}x}\right)$ and $\Psi^{(-)}_{(\lambda,2)}=\exp\left({-\sqrt{-\lambda}x}\right).$
Hence  we have that
\begin{gather*}
\zeta^{(-)}_\lambda =\big(\ln\big(\Psi^{(-)}_{\lambda}\big)\big)'
=\displaystyle{\sqrt{-\lambda}  {c_1\exp\left({\sqrt{-\lambda}x}\right)-c_2\exp\left({-\sqrt{-\lambda}x}\right)\over c_1\exp\left({\sqrt{-\lambda}x}\right)
+c_2\exp\left({-\sqrt{-\lambda}x}\right)}},\\
 \zeta_0^{(-)} =\big(\ln\big(\Psi^{(-)}_{0}\big)\big)'=\dfrac{1}{x}.
\end{gather*}
Additionally, we have that
\begin{gather*}
\zeta^{(-)}_{(\lambda,1)}=\big(\ln\big(\Psi^{(-)}_{(\lambda,1)}\big)\big)'=\sqrt{-\lambda} \qquad \mbox{and} \qquad
 \zeta^{(-)}_{(\lambda,2)}=\big(\ln\big(\Psi^{(-)}_{(\lambda,2)}\big)\big)'=-\sqrt{-\lambda}.
 \end{gather*}
Here in the expression of the vector f\/ield~\eqref{partidav} we have that $T(x)=0$ and $N(x)=1$.
Hence the polynomial vector f\/ield~\eqref{partidav} associated to the Schr\"odinger equation  $H_-\Psi^{(-)}=\lambda\Psi^{(-)}$ with
potential $V_-=0$, is
\[
X^{-}_\lambda=\big({-}\lambda-{\zeta^{(-)2}}\big)\dfrac{\partial }{\partial \zeta^{(-)}}+\dfrac{\partial }{\partial x},
\]
and is quadratic.
Then according to Lemma \ref{lemdarp} for all $\lambda\in \Lambda\setminus\{0\}$, the vector f\/ield $X^{(-)}_\lambda$ admits the following:

Invariant  curve
\[
f^{(-)}_\lambda\big(\zeta^{(-)},x\big)=-\zeta^{(-)}+\sqrt{-\lambda}  {c_1\exp\left({\sqrt{-\lambda}x}\right)-c_2\exp\left({-\sqrt{-\lambda}x}\right)\over
 c_1\exp\left({\sqrt{-\lambda}x}\right)
+c_2\exp\left({-\sqrt{-\lambda}x}\right)},
\]
and $f^{(-)}_\lambda(\zeta^{(-)},x)\in\C\big(\zeta^{(-)},x, \exp ({\sqrt{-\lambda}x} )\big)$
with generalized cofactor
\[
K^{(-)}_\lambda\big(\zeta^{(-)},x\big)
=-\zeta^{(-)}-\sqrt{-\lambda}{c_1\exp\left({\sqrt{-\lambda}x}\right)-c_2\exp\left({-\sqrt{-\lambda}x}\right)\over
c_1\exp\left({\sqrt{-\lambda}x}\right)+c_2\exp\left({-\sqrt{-\lambda}x}\right)},
\]
and $K^{(-)}_\lambda(\zeta^{(-)},x)\in\C(\zeta^{(-)},x,\exp\left({\sqrt{-\lambda}x}\right)).$

 Generalized exponential factor
\[
F^{(-)}_\lambda\big(\zeta^{(-)},x\big)=c_1\exp\big({\sqrt{-\lambda}x}\big)+c_2\exp\big({-}\sqrt{-\lambda}x \big),
\]
with generalized cofactor
\[
L^{(-)}_\lambda(\zeta^{(-)},x)=\sqrt{-\lambda}  {c_1\exp\left({\sqrt{-\lambda}x}\right)-c_2\exp\left({-\sqrt{-\lambda}x}\right)\over
 c_1\exp\left({\sqrt{-\lambda}x}\right)+c_2\exp\left({-\sqrt{-\lambda}x}\right)}\in\C\big(\zeta^{(-)},x,\exp\big({\sqrt{-\lambda}x}\big)\big).
 \]

 Generalized Darboux integrating factor $R^{(-)}_\lambda(\zeta^{(-)},x)$ given by
\begin{gather*}
{1\over \left(c_1\exp\left({\sqrt{-\lambda}x}\right)+c_2\exp\left(\!-\sqrt{-\lambda}x \right)\right)^2
\!\left(\!-\zeta^{(-)}+\sqrt{-\lambda} \dfrac{c_1\exp\left(\!{\sqrt{-\lambda}x}\right)-c_2\exp\left({\!-\sqrt{-\lambda}x}\right)}{c_1\exp\left(\!{\sqrt{-\lambda}x}\right)
+c_2\exp\left({\!-\sqrt{-\lambda}x}\right)}\right)^2},
\end{gather*}
and  note  that $R^{(-)}_\lambda(\zeta^{(-)},x)\in\C(\zeta^{(-)},x,\exp\left({\sqrt{-\lambda}x}\right)).$

 Generalized Darboux f\/irst integral{\samepage
\begin{gather*}
I^{(-)}_\lambda\big(\zeta^{(-)},x\big)=\dfrac{-\zeta^{(-)}-\sqrt{-\lambda}}{-\zeta^{(-)}+\sqrt{-\lambda}} \exp\big({-}2\sqrt{-\lambda} x \big),
 \qquad \lambda\neq 0
\end{gather*}
and $I^{(-)}_\lambda(\zeta^{(-)},x)\in\C\big(\zeta^{(-)},x,\exp ({\sqrt{-\lambda}x} )\big)$.}

Applying the Darboux transformation $\mathrm{DT}$ for $\lambda\neq 0$ (and according to Proposition \ref{propdarp}) we have
\begin{gather*}
\mathrm{DT}\big(\Psi^{(-)}_{\lambda}\big)=\Psi^{(+)}_{\lambda}=\Psi^{(-)\prime}_\lambda-\zeta^{(-)}_0\Psi^{(-)}_\lambda\\
\hphantom{\mathrm{DT}\big(\Psi^{(-)}_{\lambda}\big)}{}
 = {{{c_1(\sqrt{-\lambda}x-1)\exp\left({\sqrt{-\lambda}x}\right)\over x}-{c_2(\sqrt{-\lambda}x+1)\exp\left({-\sqrt{-\lambda}x}\right)\over x}}},\\
        \zeta_\lambda^{(+)}=\big(\ln\big({\Psi_\lambda^{(+)}}\big)\big)'=
         {-\frac{\lambda A x^2
        +\sqrt{-\lambda}Bx-A}
        {x(\sqrt{-\lambda}Bx-A)}},
\end{gather*}
with
\begin{gather*}
A=c_1 \exp\big({\sqrt{-\lambda}x}\big)+c_2 \exp\big({-\sqrt{-\lambda}x}\big),\qquad
B=c_1\exp\big({\sqrt{-\lambda}x}\big)-c_2\exp\big({-\sqrt{-\lambda}x}\big).
\end{gather*}
Additionally we have
\begin{gather*}
\zeta^{(+)}_{(\lambda,1)} = \sqrt{-\lambda}+\left(\ln\left(\sqrt{-\lambda}-\frac{1}{x}\right)\right)'
                        =\sqrt{-\lambda}+\dfrac{1}{\sqrt{-\lambda}  x^2-x},\\
\zeta^{(+)}_{(\lambda,2)} =- \sqrt{-\lambda}+\left(\ln\left(-\sqrt{-\lambda}-\frac{1}{x}\right)\right)'
                        =-\sqrt{-\lambda}-\dfrac{1}{\sqrt{-\lambda}  x^2+x}.
\end{gather*}
We can see that for all $\lambda\in \Lambda$ the Picard--Vessiot
extensions are given by $L_0=\widetilde{L}_0=\mathbb{C}(x)$,
$L_\lambda=\widetilde{L}_\lambda=\mathbb{C}\big(x,\exp ({\sqrt{-\lambda}x} )\big) $ for
$\lambda\in\mathbb{C}^*$. In this way, we have that
$\mathrm{DGal}(L_0/F)=\mathrm{DGal}(\widetilde{L}_0/F)=e$; for
$\lambda\neq0$, we have
$\mathrm{DGal}(L_\lambda/F)=\mathrm{DGal}(\widetilde{L}_\lambda/F)=\mathbb{G}_m$,
see \cite{acthesis, acmowe}.

  Now according to Proposition \ref{propdarp} we can compute $\mathrm{DT}(X^{-}_\lambda)={X}^{+}_\lambda$ and we obtain
the rational vector f\/ield
\[
{X}^+_\lambda=\left(\frac{2}{x^2}-\lambda-{\zeta^{(+)}}^2\right)\dfrac{\partial }{\partial \zeta^{(+)}}+\dfrac{\partial }{\partial x},
\qquad \forall\, \lambda\neq 0,
\]
and equivalently we  can consider the polynomial vector f\/ield  of degree four
\[
\overline{X}^+_\lambda=\big({2}-\lambda x^2-x^2{\zeta^{(+)2}} \big)\dfrac{\partial }{\partial \zeta^{(+)}}+x^2\dfrac{\partial }{\partial x},
\qquad \forall \, \lambda\neq 0,
\]
and admits the following:

  Invariant curve
\begin{gather*}
 \overline{f}_\lambda^{(+)}\big({\zeta^{(+)}},  x\big)={{f}}_\lambda^{(+)}\big({\zeta^{(+)}}, x\big)=-{\zeta}^{(+)}+\zeta_{\lambda}^{(-)}
+{\zeta_\lambda^{(-)\prime}+\frac{1}{x^2}\over \zeta_\lambda^{(-)}-\frac{1}{x}}\\
\hphantom{\overline{f}_\lambda^{(+)}\big({\zeta^{(+)}},  x\big)}{}
 = -\zeta^{(+)}-
\dfrac{c_1A_2\exp\left({\sqrt{-\lambda} x}\right)+c_2B_2\exp\left({-\sqrt{-\lambda}  x}\right)}
{x\left(c_1A_1\exp\left({\sqrt{-\lambda} x}\right)-c_2B_1\exp\left({-\sqrt{-\lambda}  x}\right)\right) },
\end{gather*}
with
\begin{gather*}
A_1 = \sqrt{-\lambda}  x-1, \qquad  B_1 = \sqrt{-\lambda}x+1,\qquad
A_2 = \lambda x^2+A_1, \qquad B_2 = \lambda x^2-B_1,
\end{gather*}
and note that ${{f}}_\lambda^{(+)}({\zeta^{(+)}},x)\in\C\big(\zeta^{(+)},x,\exp ({\sqrt{-\lambda}x} )\big)$
with generalized cofactor{\samepage
\begin{gather*}
  \overline{K}_\lambda^{(+)}\big({\zeta}^{(+)},  x\big)=x^2{{K}}_\lambda^{(+)}\big({\zeta}^{(+)},  x\big)=
x^2\left(-{\zeta}^{(+)}-\zeta_{\lambda}^{(-)}-{\zeta_\lambda^{(-)\prime}+\frac{1}{x^2}\over \zeta_\lambda^{(-)}-\frac{1}{x}}\right)\\
\hphantom{\overline{K}_\lambda^{(+)}\big({\zeta}^{(+)},  x\big)}{}
= x^2\left(-\zeta^{(+)}+\dfrac{c_1A_2\exp\left({\sqrt{-\lambda}  x}\right)+c_2B_2\exp\left({-\sqrt{-\lambda} x}\right)}
{x\left(c_1 A_1\exp\left({\sqrt{-\lambda} x}\right)-c_2B_1\exp\left({-\sqrt{-\lambda}  x}\right)\right)}\right),
 \end{gather*}
and note that $K_\lambda^{(+)}\in\C\big(\zeta^{(+)},x,\exp ({\sqrt{-\lambda}x} )\big)$.}

Generalized exponential factor
\begin{gather*}
\overline{F}_\lambda^{(+)}\big({\zeta}^{(+)},x\big) = {F}_\lambda^{(+)}\big({\zeta}^{(+)},x\big)=
{\big(\zeta_{\lambda}^{(-)}-\zeta_{0}^{(-)}\big)}\exp\left({\displaystyle{\int \zeta_{\lambda}^{(-)}}}\right)
\\
\hphantom{\overline{F}_\lambda^{(+)}\big({\zeta}^{(+)},x\big)}{}
=\dfrac{c_1A_1\exp\left({\sqrt{-\lambda}  x}\right)+c_2B_2\exp\left({-\sqrt{-\lambda}  x}\right)}{x},
\end{gather*}
with generalized cofactor
\begin{gather*}
\overline{L}_\lambda^{(+)} = x^2{L}_\lambda^{(+)}=x^2\zeta_{\lambda}^{(+)}=
x^2\left(\zeta_{\lambda}^{(-)}+{\zeta_\lambda^{(-)\prime}+\frac{1}{x^2}\over \zeta_\lambda^{(-)}-\frac{1}{x}}\right)\\
\hphantom{\overline{L}_\lambda^{(+)}}{}
 =  -x^3\dfrac{c_1A_2\exp\left({\sqrt{-\lambda}x}\right)+c_2B_2\exp\left({-\sqrt{-\lambda}}\right)}
{c_1A_1\exp\left({\sqrt{-\lambda} x}\right)+c_2B_2\exp\left({-\sqrt{-\lambda} x}\right)},
\end{gather*}
and note that
${L}_\lambda^{(+)}\in\C\big(\zeta^{(+)},x,\exp ({\sqrt{-\lambda}x} )\big)$.

Generalized Darboux integrating factor
\begin{gather*}
{\bar{R}}_\lambda^{(+)}\big({\zeta^{(+)}},x\big) = \dfrac{1}{x^2}{{R}}_\lambda^{(+)}\big({\zeta^{(+)}},x\big)
 = \dfrac{\left(c_1\exp\left({\sqrt{-\lambda}x}\right)+c_2\exp\left({-\sqrt{-\lambda}x}\right)\right)^{-2}}{x^2\left(\zeta_\lambda^{(-)}
-\frac{1}{x}\right)^2\left(-{\zeta^{(+)}}+\zeta_\lambda^{(-)}+{\zeta_\lambda^{(-)\prime}
+\frac{1}{x^2}\over \zeta_\lambda^{(-)}-\frac{1}{x}}\right)^2},
\end{gather*}
and ${\bar{R}}_\lambda^{(+)}({\zeta^{(+)}},x)\in\C\big(\zeta^{(+)},x,\exp ({\sqrt{-\lambda}x} )\big)$.

 Generalized Darboux f\/irst integral
\begin{gather*}
{\bar{I}}_\lambda^{(+)}\big(\zeta^{(+)},x\big) = {{I}}_\lambda^{(+)}\big(\zeta^{(+)},x\big)
 =
\frac{-\zeta^{(+)}-\frac{1}{xB_1}-\sqrt{-\lambda}}
{-\zeta^{(+)}+\frac{1}{xA_1}+\sqrt{-\lambda}}\left( \frac{B_1}{A_1} \right)
\exp\big({-}2\sqrt{-\lambda}  x \big),
\end{gather*}
and ${\bar{I}}_\lambda^{(+)}(\zeta^{(+)},x)\in \C\big(\zeta^{(+)},x,\exp ({\sqrt{-\lambda}x} )\big)$.

We could consider
$\zeta^{(-)}_{(\lambda,1)}=-\sqrt{\lambda}\tan \sqrt{\lambda}x$, being
$\Psi_{(\lambda,1)}^{(-)}=\cos{\sqrt{\lambda}x}$ and  $\Psi_0^{(-)}=x$. Thus we
obtain the {\it{Schr\"odinger}} polynomial vector f\/ield
\[
X^{(-)}_{(\lambda,1)}=\big({-}\lambda-\zeta^{(-)2}\big)\frac{\partial}{\partial
\zeta^{(-)}}+\frac{\partial }{\partial x},
\]
 of degree two. According to Lemma \ref{lemdarp} for all $\lambda\in \Lambda\setminus\{0\}$, the vector f\/ield $X^{(-)}_{(\lambda,1)}$
admits the following:

 Invariant  curve
\[
f^{(-)}_{(\lambda,1)}\big(\zeta^{(-)},x\big)=-\zeta^{(-)}-\sqrt{\lambda}\tan\big(\sqrt{\lambda}x\big)\in\C\big(\zeta^{(-)}, x, \tan\big(\sqrt{\lambda}x\big)\big),
\]
with generalized cofactor
\[
K^{(-)}_{(\lambda,1)}\big(\zeta^{(-)},x\big)
=-\zeta^{(-)}+\sqrt{\lambda}\tan\big(\sqrt{\lambda}x\big)\in\C\big(\zeta^{(-)}, x, \tan\big(\sqrt{\lambda}x\big)\big).
\]

Generalized exponential factor
\[
F^{(-)}_{(\lambda,1)}\big(\zeta^{(-)},x\big)=\exp\left({-\displaystyle{\sqrt{\lambda}\int \tan (\sqrt{\lambda} x)} dx}\right)
=\frac{1}{\sqrt{1+\tan^2(\sqrt{\lambda} x)}},
\]
with generalized cofactor
$L^{(-)}_{(\lambda,1)}(\zeta^{(-)},x)=-\sqrt{\lambda}\tan(\sqrt{\lambda}x)\in\C\big(\zeta^{(-)}, x, \tan(\sqrt{\lambda}x)\big)$.

Generalized Darboux integrating factor (for $\lambda\neq 0$)
\[
R^{(-)}_{(\lambda,1)}\big(\zeta^{(-)},x\big)=
\dfrac{1+\tan^2(\sqrt{\lambda} x)}{\left(-\zeta^{(-)}-\sqrt{\lambda}  \tan(\sqrt{\lambda} x)\right)^2}\in\C\big(\zeta^{(-)},x, \tan\big(\sqrt{\lambda}x\big)\big).
\]

Generalized Darboux f\/irst integral
\[
I^{(-)}_{(\lambda,1)}\big(\zeta^{(-)},x\big)=
\dfrac{\zeta^{(-)}\tan(\sqrt{\lambda}  x)-\sqrt{\lambda}}{(\zeta^{(-)}+\sqrt{\lambda} \tan(\sqrt(\lambda) x))\sqrt{\lambda}}\in\C\big(\zeta^{(-)}, x,
\tan\big(\sqrt{\lambda}x\big)\big).
\]

By Proposition \ref{propdarp}, Darboux transformation of the vector f\/ield $X^{-}_{(\lambda,1)}$ is given by
\[
{X}^{+}_{(\lambda,1)}=\left(\dfrac{2}{x^2}-\lambda -\zeta^{(+)2}\right)\frac{\partial}{\partial
\zeta^{(+)}}+\frac{\partial }{\partial x},
\]
and we can work with the polynomial vector f\/ield of degree four
\[
{\bar{X}}^{+}_{(\lambda,1)}=\big(2-\lambda x^2-x^2\zeta^{(+)2}\big)\frac{\partial}{\partial
\zeta^{(+)}}+x^2\frac{\partial }{\partial x},
\]
that admits:

 The invariant curve
\[
\bar{f}^{+}_{(\lambda,1)}\big(\zeta^{(+)},x\big) = -\zeta^{(+)}+\dfrac{-\sqrt{\lambda} x\tan\big(\sqrt{\lambda}x\big)+\lambda x^2-1}{x\big(\sqrt{\lambda}  x\tan\big(\sqrt{\lambda}x\big)+1\big)},
\]
with generalized cofactor
\[
\bar{K}^{+}_{(\lambda,1)}\big(\zeta^{(+)}, x\big)= x^2\left(-\zeta^{(+)}-\dfrac{-\sqrt{\lambda}  x\tan\big(\sqrt{\lambda}  x\big)+\lambda x^2-1}{x\big(\sqrt{\lambda}  x\tan\big(\sqrt{\lambda}x\big)+1\big)}\right).
\]

The generalized exponential factor
\[
\bar{F}^{+}_{(\lambda,1)}(\zeta^{(+)}, x) = \dfrac{-\sqrt{\lambda}\tan\big(\sqrt{\lambda}x\big)-\frac{1}{x}}{\sqrt{1+\tan^2\big(\sqrt{\lambda}  x}\big)},
\]
with generalized cofactor
\[
\bar{L}^{+}_{(\lambda,1)}\big(\zeta^{(+)}, x\big)= x\dfrac{-\sqrt{\lambda}  x\tan\big(\sqrt{\lambda}x\big)+\lambda x^2-1}{\sqrt{\lambda}  x\tan\big(\sqrt{\lambda}x\big)+1}.
\]

The integrating factor
\[
\bar{R}^{+}_{(\lambda,1)}\big(\zeta^{(+)}, x\big)=\dfrac{1+\tan^2\big(\sqrt{\lambda}x\big)}{x^2\left(-\zeta^{(+)}+\frac{\lambda x^2-A_3}
{x A_3}\right)^2
\left(-\frac{A_3}{x}\right)^2}
\]
 with
$ A_3=\sqrt{\lambda} x\tan(\sqrt{\lambda}x)+1$.

Also we could consider the partial solution $\Psi_{(\lambda,2)}^{(-)}=\sin({\sqrt{\lambda}})$ and we could work in a  similarly way as before.
\end{example}

\begin{remark}
As in the same philosophy of the original Darboux transformation, we
can iterate it to obtain families of new potentials. Starting with
$V=0$, the following potentials can be obtained using Darboux
iteration $\mathrm{DT}_n$ (see~\cite{beev,blbose,in,KKJM1,KKJM2,MAPY,TTW}):
\begin{alignat*}{5}
& I)\quad && V_n = \frac{n(n-1)b^2}{(bx+c)^2},\qquad && II)\quad && V_n = \frac{ m^2n(n-1)(b^2-a^2)}{ (a\cosh(mx)+b\sinh(mx))^2}, & \\
& III)\quad && V_n = \frac{-4abm^2n(n-1)}{ (a\exp({mx})+b\exp({-mx}))^2},\qquad &&
  IV)\quad && V_n = \frac{m^2n(n-1)(b^2+a^2)}{ (a\cos(mx)+b\sin(mx))^2}.&
\end{alignat*}
In particular for the rational potential given in~\textit{I}), see~\cite{acthesis,acmowe}, we have $F=F_n=\mathbb{C}(x)$
and for $\lambda_n=\lambda=0$, we have
\[
\Psi_0^{(n)}={c_1\over
(bx+c)^n}+c_2(bx+c)^{n+1},
\]
 so that
\[
\mathrm{DGal}(L_0/F)=\mathrm{DGal}\big(L_0^{(n)}/F\big)=e,
\]
 whilst for $\lambda\neq 0$
and $\lambda_n=0$, the general solution $\Psi_\lambda^{(n)}$ is
given by
\[
\Psi_\lambda^{(n)}(x)=c_1A_n(x,\lambda)C_n\big(\sin\big(\sqrt{\lambda}x\big)+c_2B_n(x,\lambda)D_n\big(\cos\big(\sqrt{\lambda}x\big)\big)\big),
\]
where $A_n,B_n,C_n,D_n\in\mathbb{C}(x)$, so that
\[
\mathrm{DGal}(L_\lambda/F)=\mathrm{DGal}\big(L^{(n)}_\lambda/F\big)=\mathbb{G}_m.
  \]
  Finally, considering $b=1$, $n=\ell+1$, $c=0$ and $x=r$ we arrive to the \emph{square well potential}, which can be considered in a similar way.
\end{remark}

\begin{example}[3D Harmonic oscillator]
We take $V_{-}=r^2+{\ell(\ell +1)\over r^2}-(2\ell +3)$, following \cite{acthesis,acmowe} we see that $\Lambda\cap\mathrm{Spec}(H)=4\mathbb{N}$. For instance we consider $\lambda=4n$, where $n\in\mathbb{N}$.
Additionally, we have
\begin{gather*}
\Psi^{(-)}_0=r^{\ell +1}\exp\left({-\frac{r^2}{2}}\right),  \qquad
\zeta^{(-)}_0={\ell+1\over r}-r\in\mathbb{C}(r),\qquad
\Psi^{(-)}_n=r^{\ell+1}P_n\exp\left({-\frac{r^2}{2}}\right),\\ \zeta_{(n,1)}^{(-)}=\dfrac{\ell+1}{r}+\dfrac{P'_n}{P_n}-r,\qquad
\zeta_{(n,2)}^{(-)}=\dfrac{\exp({r^2})}{r^{2\ell+2}P_n^2}\displaystyle{\int}\dfrac{\exp\left({r^2}\right)}{r^{2\ell+2}P_n^2}dx+\dfrac{P_n'}{P_n}
-r+\dfrac{\ell+1}{r},
\end{gather*}
where $P_n$ are polynomials of degree $n$ which are related to the \emph{generalized Laguerre polynomials}, see~\cite{niuv}.  According to Proposition~\ref{darbiso} we see that $\mathrm{DT}$ is strong
isogaloisian. Since,
\[
\mathrm{DT}(V_-)=V_+=r^2+{(\ell+1)(\ell +2)\over r^2}-(2\ell +1),
\]
we have that the potential $V_-$ is also shape invariant.

Note that $T(r)=r^4+{\ell(\ell +1)}-(2\ell +3)r^2$ and
$N(r)=r^2$. Hence,  system \eqref{partidav} can be written as
\[
X^{-}_n=\big(r^4+\ell(\ell +1)-(2\ell +3)r^2-4nr^2- r^2{\zeta^{(-)2}}\big)\dfrac{\partial}{ \partial {\zeta^{(-)}}}
+r^2\dfrac{\partial}{ \partial r}.
\]
According to Lemma \ref{lemdarp} admits the invariant curve
\[
f^{(-)}_n\big(\zeta^{(-)},r\big)=-\zeta^{(-)}+{\ell+1\over r}+\frac{P_n'}{P_n}-r\in\mathbb{C}\big(\zeta^{(-)},r\big),
\]
with generalized cofactor
\[
 K^{(-)}_n\big(\zeta^{(-)},r\big)=-r^2\zeta^{(-)}-({\ell+1})r-\frac{P_n'}{P_n}r+r^2\in\mathbb{C}\big(\zeta^{(-)},r\big).
\]

The generalized exponential factor
\[
F^{(-)}_n\big(\zeta^{(-)},r\big)=r^{\ell+2}P_n\exp\left({-r^2\over 2}\right)\in L^2,
\]
with generalized cofactor
\[
 L^{(-)}_n\big(\zeta^{(-)},r\big)=r+({\ell+1})r+\frac{P_n'}{P_n}r^2-r^3\in\mathbb{C}(r).
\]
 Moreover, $X^{-}_\lambda$ admits the generalized Darboux integrating factor
\[
R^{(-)}_n\big(\zeta^{(-)},r\big)=\dfrac{\exp\left({r^2}\right)}{P_n^2r^{2(\ell+2)}
\left(-\zeta^{(-)}+{\ell+1\over r}+\frac{P_n'}{P_n}-r\right)^2},
\]
where $R^{(-)}_\lambda(\zeta^{(-)},r)\in\C(r,\exp\left({r^2/2}\right))$
 and the f\/irst integral
\[
I_n^{(-)}\big(\zeta^{(-)},r\big)=\dfrac{\exp\left({r^2}\right)}{r^{2\ell+2}P_nP_n'-r^{2\ell+1}P_n^2A_1}
+\displaystyle{\int\dfrac{\exp\left({r^2}\right)}{r^{2\ell+1}P_n^2}dr}\in F,
\]
with $A_1=r^2+\zeta^{(-)}r-\ell-1.$
According to Proposition \ref{propdarp} using the  Darboux transformation we obtain the rational vector f\/ield
\[
X^{+}_n=\left(r^2+{(\ell+1)(\ell +2)\over r^2}-(2\ell +1)-4n-{\zeta^{(+)2}}\right)\dfrac{\partial}{\partial \zeta^{(+)}}
      +\dfrac{\partial
}{\partial r},
\]
or equivalently we can consider the polynomial vector f\/ield
\[
\bar{X}^{+}_\lambda= \big(r^4+{(\ell+1)(\ell +2)}-(2\ell +1)r^2-4n r^2-r^2{\zeta^{(+)2}} \big)\frac{\partial}{\partial \zeta^{(+)}}
+r^2\dfrac{\partial}{\partial r}.
\]
Moreover, we have
\begin{gather*}
\zeta_{(n,1)}^{(+)} = \frac{P_n''}{P_n'}-r+\frac{4n+1}{r},\\
\zeta_{(n,2)}^{(+)} = r^{2\ell+2}P_n\big(rP_n''-\big(r^2-\ell-1\big)P_n'\big)
 \int \frac{\exp ({r^2} )}{r^{2\ell+2}P_n^2}+\exp\big({r^2}\big)\big(r^2-\ell-1\big).
\end{gather*}
The vector f\/ield ${\bar{X}}^{+}_n$ admits the invariant curve
\[
\bar{f}^{(+)}_n\big(\zeta^{(+)},r\big)=-\zeta^{(+)}+{\ell+1\over r}+\frac{P''_n}{P'_n}-r\in\mathbb{C}\big(\zeta^{(+)},r\big),
\]
with generalized cofactor
\[
\bar{K}^{(+)}_n\big(\zeta^{(+)},r\big)=-r^2\zeta^{(+)}-r({\ell+1})-\frac{P_n''}{P'_n}r^2+r^3\in\mathbb{C}\big(\zeta^{(+)},r\big).
\]
The generalized exponential factor
\[
 \bar{F}^{(+)}_n\big(\zeta^{(+)},r\big)=r^{\ell+2}P'_n\exp\left( - {r^2\over 2}\right)\in L^2,
\]
with generalized cofactor
\[
 \bar{L}^{(+)}_n\big(\zeta^{(+)},r\big)=r+({\ell+1})r-\frac{P_n''}{P'_n}r^2+r^3\in\mathbb{C}(r).
\]
Additionally, the vector f\/ield $\bar{X}^{+}_n$  admits the generalized Darboux integrating factor
\[
 R^{(+)}_n\big(\zeta^{(+)},r\big)=\frac{\exp\left({r^2}\right)}{{P'}^2_nr^{2(\ell+2)}\left(-\zeta^{(+)}+\frac{\ell+1}{ r}+\frac{P''_n}
{P'_n}-r\right)^2}\in\C\big(r,\exp\big({r^2/2}\big)\big),
\]
and the f\/irst integral
\[
I_n^{(+)}\big(\zeta^{(+)},r\big)=\dfrac{P_n'P_nr^{2\ell+2}\left(rP_n''-B_2P_n'\right)
I_1+A}
{\left(rP_n''-B_2P_n'\right)\left(\exp\left({r^2}\right)+r^{2\ell+2}(P_nP_n')^2I_1\right)},
\]
with
\begin{gather*}
I_1 = \displaystyle{\int}\dfrac{\exp ({r^2} )}{r^{2\ell+2}P_n^2}dr,\qquad
A = P_n'\exp\big({r^2}\big)B_1\exp\left({\displaystyle{\int}\Phi}\right),\qquad
B_1 = r^2-\zeta^{(+)}r-\ell-1,\\
B_2=r^2+\zeta^{(+)}r-\ell-1,\qquad
\Phi = \frac{\exp ({r^2} ) ((2r^2-2\ell-2)P_n'-rP_n'' )}
{rP_n'\left(\exp\left({r^2}\right)+r^{2\ell+2}P_nP_n' {\int}\frac{\exp ({r^2} )}{r^{2\ell+2}P_n^2}dr\right)}.
\end{gather*}
Note that $I_n^{(\pm)}(\zeta^{(\pm)},r)\in F$, being $F$ an extension of $\mathbb{C}(r,\zeta^{(\pm)})$ adding the solutions of the Schr\"odinger equation $H^\pm\Psi=\lambda\Psi$.
\end{example}

\begin{example}[Coulomb]
In this case we have $V_{-}={\ell(\ell +1)\over r^2}-{2(\ell +1)\over r}+1$ and we can take
\[
\Psi^{(-)}_0=r^{\ell +1}\exp(-r),  \qquad \zeta^{(-)}_0={\ell+1\over r}-1\in\mathbb{C}(r).
\]
Note that due to Proposition \ref{darbiso} we have that $\mathrm{DT}$ is strong isogaloisian. Since
\[
\mathrm{DT}(V_-)=V_+={(\ell+1)(\ell +2)\over r^2}-{2(\ell +1)\over r}+1,
\]
we have that the potential $V_-$ is shape invariant.

It is hold that
\[
\Lambda=\left\{1-\left({\ell+1\over\ell+1+n}\right)^2:
n\in\mathbb{Z}_+\right\}\cup
\left\{1-\left({\ell+1\over\ell-n}\right)^2:
n\in\mathbb{Z}_+\right\},
\] see \cite{acthesis,acmowe}. In particular, for $\lambda=1-\left({\ell+1\over\ell+1+n}\right)^2$, $\lambda\neq 0$, we have
\begin{gather*}
\Psi^{(-)}_n=r^{\ell+1}P_n \exp\left({\frac{-(\ell+1)r}{\ell+1+n}}\right), \qquad
\zeta^{(-)}_{(n,1)}={\ell+1\over r}+\frac{P'_n}{P_n}-{\ell+1\over \ell +1+n}\in\mathbb{C}(r),\\
\zeta_{(n,2)}^{(+)}=\dfrac{\ell+1}{r}+\dfrac{P'_n}{P_n}-\dfrac{\ell+1}{\ell+n+1}-1+\dfrac{\exp\left(\frac{4r-2nr}{\ell+n+1}\right)}{r^{2\ell+2}P_n^2
\int\frac{\exp\left(\frac{4r-2nr}{\ell+n+1}\right)}{r^{\ell+2}P_n^2}},
\end{gather*}
where $P_n$ are polynomials of degree $n$, which are related to the \emph{generalized Laguerre polyno\-mials}~$L^{(\ell)}_n$, for more details see~\cite{niuv}.

According to Lemma~\ref{lemdarp} for $T(r)={\ell(\ell +1)}-{2(\ell +1)r}+r^2$ and $N(r)=r^2$  the vector f\/ield~\eqref{partidav} becomes
\[
X^{(-)}_n=\left({\ell(\ell +1)}-{2(\ell +1)r}+\left({\ell+1\over\ell+1+n}\right)^2r^2- r^2{\zeta^{(-)2}}\right){\partial\over \partial {\zeta^{(-)}}}
+r^2{\partial\over \partial r},
\]
and admits:

The invariant curve
\[
  f^{(-)}_n\big(\zeta^{(-)},r\big)=-\zeta^{(-)}+{\ell+1\over r}+\frac{P'_n}{P_n}-{\ell+1\over \ell +1+n}\in\mathbb{C}\big(\zeta^{(-)},r\big),
\]
with generalized cofactor
\[
 K^{(-)}_n\big(\zeta^{(-)},r\big)=-r^2\zeta^{(-)}-{(\ell+1)r}-\frac{P'_n}{P_n}r^2+{\ell+1\over \ell +1+n}r^2\in\mathbb{C}\big(\zeta^{(-)},r\big).
\]

 The  exponential factor
\[
F^{(-)}_n\big(\zeta^{(-)},r\big)=r^{\ell+2}P_n \exp\left({\frac{-(\ell+1)r}{\ell+1+n}}\right)\in L^2,
\]
with generalized cofactor
\[
L^{(-)}_n\big(\zeta^{(-)},r\big)=r+{(\ell+1)r}+\frac{P'_n}{P_n}r^2-{\ell+1\over \ell +1+n}r^2\in\mathbb{C}(r).
\]

 We note that the vector f\/ield $X^{-}_n$   admits
 the generalized Darboux integrating factor
\begin{gather*}
R^{(-)}_n\big(\zeta^{(-)},r\big)=\frac{\exp\left({2(\ell+1)r\over \ell+1+n}\right)}{P_{n}^2r^{2(\ell+2)}\!\left(-\zeta^{(-)}\!+{\ell+1\over r}
+\frac{P'_n}{P_n}-{\ell+1\over \ell +1+n}\right)^2}\in\C\left(\zeta^{(-)},r,\exp\left(\frac{(\ell+1)r}{\ell+1+n}\right)\right).
\end{gather*}

The f\/irst integral
\begin{gather*}
I_n^{(-)}(\zeta^{(-)}, r)=\left(\frac{(\ell+n+1)r^{-2\ell-1}\exp\left({\frac{4r-2nr}{\ell+n+1}}\right)}
{P_n\left(
(\ell +n+1)rP_n'+(\ell+1)(\ell+n+1)-AI_1
\right)}+1\right) \exp\left({\displaystyle{\int} \phi}\right)\in F,
\end{gather*}
with
\begin{gather*}
A = \ell r(\zeta^{(-)}+2)+nr(\zeta^{(-)}+1)+r(\zeta^{(-)}+2)P_n,\\
I_1=\displaystyle{\int}\dfrac{\exp\left({\frac{4r-2nr}{\ell+n+1}}\right)}{r^{2\ell+2}P_n^2}dr,\qquad
\Phi=\dfrac{\exp\left({\frac{4r-2nr}{\ell+n+1}}\right)}{r^{2\ell+2}P_n^2\displaystyle{\int} \dfrac{\exp\left({\frac{4r-2nr}{\ell+n+1}}\right)}{r^{2\ell+2}P_n^2}}.
\end{gather*}

Note that
\begin{gather*}
\zeta_{(n,1)}^{(+)} = \dfrac{\ell+1}{r}-\dfrac{3\ell+n+3}{\ell+n+1}+\dfrac{(\ell+n+1)P_n''}{((\ell+n+1)P_n'-(\ell+1)P_n)}\\
\hphantom{\zeta_{(n,1)}^{(+)} =}{}
-\dfrac{(\ell+1)^2P_n}{(\ell+n+1)((\ell+n+1)P_n'-(\ell+1)P_n)}.
\end{gather*}
 For suitability let assume $\mu=\ell+n+1$, thus we also have
\[
\zeta_{(n,2)}^{(+)}=\dfrac{\exp\left({\frac{2nr}{\mu}}\right)r^{2\ell+2}P_n\chi A_1P_n''+B_1P_n+\exp\left({4r}\right)\mu^2(r-\ell-1)}
{r\mu\left(\exp\left({\frac{2nr}{\mu}}\right)r^{2\ell+2}P_n\chi(\mu P_n'-(\ell+1)P_n)+\mu \exp({4r})\right)},
\]
with
\begin{gather*}
A_1 = r\mu^2P_n''+\mu((\ell+1)\mu-r(3\ell+n+3)),\qquad
B_1=(\ell+1)(r(2\ell+n+2)-(\ell+1)\mu),\\
\chi=\displaystyle{\int}\dfrac{\exp\left(\frac{4r-2nr}{\mu}\right)}{r^{2\ell+2}P_n^2}dr.
\end{gather*}
After the Darboux transformation $\mathrm{DT}$ the vector f\/ield $X^{-}_n$ becomes
\[
X^{(+)}_n=\left({(\ell+1)(\ell +2)\over r^2}-{2(\ell +1)\over r}+1-\lambda-{\zeta^{(+)2}}\right)\frac{\partial}{\partial \zeta^{(+)}}
 +\frac{\partial}{\partial r}.
\]
The vector f\/ield ${X}^{+}_n$ admits
the invariant curve
\[
f^{(+)}_n\big(\zeta^{(+)},r\big)=-\zeta^{(+)}+{\ell+1\over r}+\frac{P''_n}{P'_n}-{\ell+1\over \ell +1+n}\in\mathbb{C}\big(\zeta^{(+)},r\big),
\]
 with generalized cofactor
\[
 K^{(+)}_n\big(\zeta^{(+)},r\big)=-\zeta^{(+)}-{\ell+1\over r}-\frac{P''_n}{P'_n}+{\ell+1\over \ell +1+n}\in\mathbb{C}\big(\zeta^{(+)},r\big).
\]
 The generalized exponential factor
\[
 F^{(+)}_n\big(\zeta^{(+)},r\big)=r^{\ell+1}P'_n \exp\left({\frac{-(\ell+1)r}{\ell+1+n}}\right)\in L^2,
\]
 with generalized cofactor
\[
L^{(+)}_n\big(\zeta^{(+)},r\big)={\ell+1\over r}+\frac{P''_n}{P'_n}-{\ell+1\over \ell +1+n}\in\mathbb{C}(r).
\]
Hence, the vector f\/ield ${X}^{+}_n$ admits
the generalized Darboux integrating factor
\[
R^{(+)}_n\big(\zeta^{(+)},r\big)=\dfrac{\exp\left({2(\ell+1)r\over \ell+1+n}\right)}{P_{n}^2r^{2(\ell+1)}
\left(-\zeta^{(+)}+{\ell+1\over r}+\frac{P''_n}{P'_n}-{\ell+1\over \ell +1+n}\right)^2},
\]
and
\[
R^{(+)}_n(\zeta^{(+)},r)\in\C\left(\zeta^{(+)},r,\exp\left({\frac{(\ell+1)r}{\ell+1+n}}\right)\right).
\]
The f\/irst integral is given by relation
\[
I\big(\zeta^{(+)},r\big)=\dfrac{-\zeta^{(+)}+\zeta_{(\lambda,2)}^{(+)}}{-\zeta^{(+)}+\zeta_{(\lambda,1)}^{(+)}}
\exp\left({\displaystyle{\int}\big(\zeta_{(\lambda,2)}^{(+)}-\zeta_{(\lambda,1)}^{(+)}\big)dr}\right)\in F.
\]
Note that $I_n^{(\pm)}(\zeta^{(\pm)},r)\in F$, being $F$ an extension of $\mathbb{C}(r,\zeta^{(\pm)})$ adding the solutions of the Schr\"odinger equation $H^\pm\Psi=\lambda\Psi$.
\end{example}

\section{Final remarks}

\looseness=-1
This work is a f\/irst approach of applying Darboux transformation
into polynomial vector f\/ields that its associated foliation is of
Riccati type. The strong isogaloisian property of the Darboux
transformation guarantees that the transformed vector f\/ield is also
of Riccati type. Additional\-ly, we show that in the case of  shape
invariant potential is preserved the rational structure of the
invariant objects such as the invariant curves, generalized
cofactors, generalized Darboux integrating factors and Darboux f\/irst
integrals. We remark that in this work we analyze the case of
algebraically solvable potentials and in particular we consider
potentials satisfying the shape invariance condition. A natural
question arises: What happen in the case of quasi-algebraically
solvable potentials? A more general study should be also done  in
the future for non rational potentials.

\subsection*{Acknowledgements}

The authors are  partial supported by the MICINN/FEDER grant number MTM2009-06973. The f\/irst author is also supported by Research Department
of Universidad del Norte grant Agenda 2012.
CP is additionally partially supported by the MICINN/FEDER grant MTM2008--03437 and  by the Generalitat de Catalunya grant number 2009SGR859.
The authors acknowledge to the anonymous referees by their useful comments and suggestions.

\pdfbookmark[1]{References}{ref}
\LastPageEnding


\begin{thebibliography}{99}
\footnotesize\itemsep=0.5pt

\bibitem{acthesis}
Acosta-Hum{\'a}nez P.B., Galoisian approach to supersymmetric quantum
  mechanics. The integrability analysis of the Schr\"odinger equation by means
  of dif\/ferential Galois theory, VDM Verlag, Dr M\"uller, Berlin, 2010.

\bibitem{almp}
Acosta-Hum\'anez P.B., L\'azaro-Ochoa J.T., Morales-Ruiz J.J., Pantazi Ch., On
  the integrability of polynomial f\/ields in the plane by means of
  Picard--Vessiot theory, \href{http://arxiv.org/abs/1012.4796}{arXiv:1012.4796}.

\bibitem{acmowe}
Acosta-Hum{\'a}nez P.B., Morales-Ruiz J.J., Weil J.A., Galoisian approach to
  integrability of {S}chr\"odinger equation, \href{http://dx.doi.org/10.1016/S0034-4877(11)60019-0}{\textit{Rep. Math. Phys.}}
  \textbf{67} (2011), 305--374, \href{http://arxiv.org/abs/1008.3445}{arXiv:1008.3445}.

\bibitem{beev}
Berkovich L.M., Evlakhov S.A., The {E}uler--{I}mshenetski\u\i--{D}arboux
  transformation of second-order linear equations, \href{http://dx.doi.org/10.1134/S0361768806030066}{\textit{Program. Comput.
  Software}} \textbf{32} (2006), 154--165.

\bibitem{BP1}
Bl\'azquez-Sanz D., Pantazi Ch., A~note on the Darboux theory of integrability
  of non autonomous polynomial dif\/ferential systems, Preprint, 2011.

\bibitem{blbose}
Blecua P., Boya L.J., Segui A., New solvable quantum-mechanical potentials by
  iteration of the free {$V=0$} potential, \href{http://dx.doi.org/10.1393/ncb/i2002-10008-y}{\textit{Nuovo Cimento~B}}
  \textbf{118} (2003), 535--546, \href{http://arxiv.org/abs/quant-ph/0311139}{quant-ph/0311139}.

\bibitem{Car}
Carnicer M.M., The {P}oincar\'e problem in the nondicritical case, \href{http://dx.doi.org/10.2307/2118601}{\textit{Ann.
  of Math.~(2)}} \textbf{140} (1994), 289--294.

\bibitem{CeLi}
Cerveau D., Lins~Neto A., Holomorphic foliations in {${\bf C}{\rm P}(2)$}
  having an invariant algebraic curve, \textit{Ann. Inst. Fourier (Grenoble)}
  \textbf{41} (1991), 883--903.

\bibitem{CL1}
Christopher C., Llibre J., Algebraic aspects of integrability for polynomial
  systems, \href{http://dx.doi.org/10.1007/BF02969405}{\textit{Qual. Theory Dyn. Syst.}} \textbf{1} (1999), 71--95.

\bibitem{CLPW2}
Christopher C., Llibre J., Pantazi Ch., Walcher S., Inverse problems for
  invariant algebraic curves: explicit computations, \href{http://dx.doi.org/10.1017/S0308210507001175}{\textit{Proc. Roy. Soc.
  Edinburgh Sect.~A}} \textbf{139} (2009), 287--302.


\bibitem{CLPW}
Christopher C., Llibre J., Pantazi Ch., Walcher S., Inverse problems for
  multiple invariant curves, \href{http://dx.doi.org/10.1017/S0308210506000400}{\textit{Proc. Roy. Soc. Edinburgh Sect.~A}}
  \textbf{137} (2007), 1197--1226.


\bibitem{CLPZ}
Christopher C., Llibre J., Pantazi Ch., Zhang X., Darboux integrability and
  invariant algebraic curves for planar polynomial systems, \href{http://dx.doi.org/10.1088/0305-4470/35/10/310}{\textit{J.~Phys.~A:
  Math. Gen.}} \textbf{35} (2002), 2457--2476.

\bibitem{CLP}
Christopher C., Llibre J., Pereira J.V., Multiplicity of invariant algebraic
  curves in polynomial vector f\/ields, \href{http://dx.doi.org/10.2140/pjm.2007.229.63}{\textit{Pacific~J. Math.}} \textbf{229}
  (2007), 63--117.

\bibitem{cokasu2}
Cooper F., Khare A., Sukhatme U., Supersymmetry in quantum mechanics, \href{http://dx.doi.org/10.1142/9789812386502}{World
Scientif\/ic Publishing Co. Inc.}, River Edge, NJ, 2001.

\bibitem{Da}
Darboux G., M\'emoire sur les \'equations dif\/f\'erentielles alg\'ebriques du
  premier ordre et du premier degr\'e, \textit{Bull. Sci. Math.~(2)} \textbf{2}
  (1878), 60--96, 123--144, 151--200.

\bibitem{da1}
Darboux G., Sur une proposition relative aux \'equations lin\'eaires,
  \textit{Comptes Rendus Acad. Sci.} \textbf{94} (1882), 1456--1459.

\bibitem{da2}
Darboux G., Th\'eorie des Surfaces,~II, Gauthier-Villars, Paris, 1889.

\bibitem{GGG}
Garc{\'{\i}}a I.A., Giacomini H., Gin{\'e} J., Generalized nonlinear
  superposition principles for polynomial planar vector f\/ields, \textit{J.~Lie
  Theory} \textbf{15} (2005), 89--104.

\bibitem{GG}
Garc{\'{\i}}a I.A., Gin{\'e} J., Generalized cofactors and nonlinear
  superposition principles, \href{http://dx.doi.org/10.1016/S0893-9659(03)90107-8}{\textit{Appl. Math. Lett.}} \textbf{16} (2003),
  1137--1141.

\bibitem{ge}
Gendenshte\"{\i}n L.E., Derivation of exact spectra of the Schr\"odinger
  equation by means of supersymmetry, \textit{JETP Lett.} \textbf{38} (1983),
  356--359.

\bibitem{GL}
Gin{\'e} J., Llibre J., A family of isochronous foci with {D}arboux f\/irst
  integral, \href{http://dx.doi.org/10.2140/pjm.2005.218.343}{\textit{Pacific~J. Math.}} \textbf{218} (2005), 343--355.

\bibitem{in}
Ince E.L., Ordinary dif\/ferential equations, Dover Publications, New York, 1944.

\bibitem{Jo}
Jouanolou J.P., \'{E}quations de {P}faf\/f alg\'ebriques, \textit{Lecture Notes
  in Mathematics}, Vol.~708, Springer, Berlin, 1979.

\bibitem{KKJM1}
Kalnins E.G., Kress J.M., Miller W., Families of classical subgroup
  separable superintegrable systems, \href{http://dx.doi.org/10.1088/1751-8113/43/9/092001}{\textit{J.~Phys.~A: Math. Theor.}}
  \textbf{43} (2010), 092001, 8~pages, \href{http://arxiv.org/abs/0912.3158}{arXiv:0912.3158}.

\bibitem{KKJM2}
Kalnins E.G., Kress J.M., Miller W., Superintegrability and higher order
  integrals for quantum systems, \href{http://dx.doi.org/10.1088/1751-8113/43/26/265205}{\textit{J.~Phys.~A: Math. Theor.}} \textbf{43}
  (2010), 265205, 21~pages, \href{http://arxiv.org/abs/1002.2665}{arXiv:1002.2665}.

\bibitem{ka}
Kaplansky I., An introduction to dif\/ferential algebra, Hermann, Paris, 1957.

\bibitem{kol}
Kolchin E.R., Dif\/ferential algebra and algebraic groups, \textit{Pure and
  Applied Mathematics}, Vol.~54, Academic Press, New York, 1973.

\bibitem{ko}
Kovacic J.J., An algorithm for solving second order linear homogeneous
  dif\/ferential equations, \href{http://dx.doi.org/10.1016/S0747-7171(86)80010-4}{\textit{J.~Symbolic Comput.}} \textbf{2} (1986),
  3--43.

\bibitem{LL}
Llibre J., On the integrability of the dif\/ferential systems in dimension two
  and of the polynomial dif\/ferential systems in arbitrary dimension,
  \textit{J.~Appl. Anal. Comput.} \textbf{1} (2011), 33--52.

\bibitem{LlPantazi}
Llibre J., Pantazi Ch., Darboux theory of integrability for a class of
  nonautonomous vector f\/ields, \href{http://dx.doi.org/10.1063/1.3205450}{\textit{J.~Math. Phys.}} \textbf{50} (2009),
  102705, 19~pages.

\bibitem{LR}
Llibre J., Rodr{\'{\i}}guez G., Conf\/igurations of limit cycles and planar
  polynomial vector f\/ields, \href{http://dx.doi.org/10.1016/j.jde.2003.10.008}{\textit{J.~Differential Equations}} \textbf{198}
  (2004), 374--380.

\bibitem{LLZ1}
Llibre J., Zhang X., Rational f\/irst integrals in the {D}arboux theory of
  integrability in {${\mathbb C}^n$}, \href{http://dx.doi.org/10.1016/j.bulsci.2007.12.001}{\textit{Bull. Sci. Math.}} \textbf{134}
  (2010), 189--195.

\bibitem{MAPY}
Maciejewski A.J., Przybylska M., Yoshida H., Necessary conditions for classical
  super-integrability of a~certain family of potentials in constant curvature
  spaces, \href{http://dx.doi.org/10.1088/1751-8113/43/38/382001}{\textit{J.~Phys.~A: Math. Theor.}} \textbf{43} (2010), 382001,
  15~pages, \href{http://arxiv.org/abs/1004.3854}{arXiv:1004.3854}.

\bibitem{niuv}
Nikiforov A.F., Uvarov V.B., Special functions of mathematical physics.
  A~unif\/ied introduction with applications, Birkh\"auser Verlag, Basel, 1988.

\bibitem{Pan}
Pantazi Ch., Inverse problems of the Darboux theory of integrability for planar
  polynomial dif\/ferential systems, Ph.D. thesis, Universitat Autonoma de
  Barcelona, 2004.

\bibitem{PS}
Prelle M.J., Singer M.F., Elementary f\/irst integrals of dif\/ferential equations,
  \href{http://dx.doi.org/10.2307/1999380}{\textit{Trans. Amer. Math. Soc.}} \textbf{279} (1983), 215--229.

\bibitem{martinetramis}
Ramis J.P., Martinet J., Th\'eorie de {G}alois dif\/f\'erentielle et resommation,
  in Computer Algebra and Dif\/ferential Equations, \textit{Comput. Math. Appl.}, Academic
  Press, London, 1990, 117--214.

\bibitem{Sc}
Schlomiuk D., Algebraic particular integrals, integrability and the problem of
  the center, \href{http://dx.doi.org/10.2307/2154430}{\textit{Trans. Amer. Math. Soc.}} \textbf{338} (1993), 799--841.

\bibitem{Si}
Singer M.F., Liouvillian f\/irst integrals of dif\/ferential equations,
  \href{http://dx.doi.org/10.2307/2154053}{\textit{Trans. Amer. Math. Soc.}} \textbf{333} (1992), 673--688.

\bibitem{spiridonov}
Spiridonov V., Universal superpositions of coherent states and self-similar
  potentials, \href{http://dx.doi.org/10.1103/PhysRevA.52.1909}{\textit{Phys. Rev.~A}} \textbf{52} (1995), 1909--1935,
  \href{http://arxiv.org/abs/quant-ph/9601030}{quant-ph/9601030}.

\bibitem{te}
Teschl G., Mathematical methods in quantum mechanics. With applications to
  Schr{\"o}dinger operators, \textit{Gra\-dua\-te Studies in Mathematics}, Vol.~99,
  American Mathematical Society, Providence, RI, 2009.

\bibitem{TTW}
Tremblay F., Turbiner A.V., Winternitz P., An inf\/inite family of solvable and
  integrable quantum systems on a plane, \href{http://dx.doi.org/10.1088/1751-8113/42/24/242001}{\textit{J.~Phys.~A: Math. Theor.}}
  \textbf{42} (2009), 242001, 10~pages, \href{http://arxiv.org/abs/0904.0738}{arXiv:0904.0738}.

\bibitem{vasi}
van~der Put M., Singer M.F., Galois theory of linear dif\/ferential equations,
  \textit{Grundlehren der Mathematischen Wissenschaften}, Vol.~328,
  Springer-Verlag, Berlin, 2003.

\bibitem{we2}
Weil J.A., Introduction to dif\/ferential algebra and dif\/ferential Galois theory,
  CIMPA-UNESCO Lectures, Hanoi, 2001.

\bibitem{wi}
Witten E., Dynamical breaking of supersymmetry, \href{http://dx.doi.org/10.1016/0550-3213(81)90006-7}{\textit{Nuclear Phys.~B}}
  \textbf{188} (1981), 513--554.

\bibitem{Zol}
{\.Z}o{\l}\c{a}dek H., Polynomial {R}iccati equations with algebraic solutions,
  in Dif\/ferential {G}alois Theory ({B}\c{e}dlewo, 2001), \href{http://dx.doi.org/10.4064/bc58-0-17}{\textit{Banach Center
  Publ.}}, Vol.~58, Polish Acad. Sci., Warsaw, 2002, 219--231.

\end{thebibliography}
\end{document}